\documentclass[11pt,twoside]{article}
\usepackage{atmp}
\copyrightnotice{2002}{6}{403}{456}	 
\input epsf.sty

\newcommand{\ra}{\rangle}
\newcommand{\la}{\langle}

\newcommand{\vp}{\varphi}

\newcommand{\B}{b'}
\newcommand{\C}{c'}
\newcommand{\bB}{\bar b'}
\newcommand{\bC}{\bar c'}
\newcommand{\Bu}{B_{\vec u}}
\newcommand{\VV}{{\cal V}}
\newcommand{\II}{{\cal I}}

\newcommand{\MM}{{\cal M}}
\newcommand{\CC}{{\cal C}}
\newcommand{\OO}{{\cal O}}
\newcommand{\QQ}{{\cal Q}}

\newcommand{\EE}{{\cal E}}

\newcommand{\rrr}{\rangle\rangle}

\newcommand{\wt}{\widetilde}
\newcommand{\wh}{\widehat}

\newcommand{\NN}{{\cal N}}
\newcommand{\SSS}{{\cal S}}
\newcommand{\TT}{{\cal T}}

\newcommand{\be}{\begin{equation}}
\newcommand{\ee}{\end{equation}}
\newcommand{\ben}{\begin{eqnarray}\displaystyle}
\newcommand{\een}{\end{eqnarray}}
\newcommand{\refb}[1]{(\ref{#1})}
\newcommand{\p}{\partial}
\newcommand{\sectiono}[1]{\section{#1}\setcounter{equation}{0}}

\begin{document}

\title{ Ghost Structure and Closed Strings in
Vacuum String Field Theory }

 \url{hep-th/0111129}

\author{Davide Gaiotto$^a$, Leonardo Rastelli$^a$,
Ashoke Sen$^b$ and Barton Zwiebach$^c$}
\address{$^a$Department of Physics 
\\ Princeton University, Princeton, NJ 08540,
USA }
\addressemail{dgaiotto@princeton.edu,
        rastelli@feynman.princeton.edu}

\address{$^b$Harish-Chandra Research
Institute\\ Chhatnag Road, Jhusi,
Allahabad 211019, INDIA}
\addressemail{asen@thwgs.cern.ch, sen@mri.ernet.in}

\address{$^c$Center for Theoretical Physics \\
Massachussetts Institute of Technology\\
Cambridge,
MA 02139, USA}
\addressemail{zwiebach@mitlns.mit.edu}


\markboth{\it GHOST STRUCTURE AND CLOSED STRINGS\ldots}{\it D.
GAIOTTO, ET. AL.}

\begin{abstract}
We complete the construction of vacuum string field theory
by proposing a canonical choice of ghost
kinetic term -- a local insertion of the ghost field at the string
midpoint with an infinite normalization.
This choice, supported by level expansion studies in the
Siegel 
gauge,  

\cutpage 

\noindent
allows a simple analytic treatment of the ghost
sector of the string field equations. As a result, solutions
are just projectors, such as the sliver, of an auxiliary
CFT built by combining the matter part with a twisted
version of the ghost conformal theory.
Level expansion experiments lead to surprising new projectors --
butterfly surface states, whose analytical expressions 
are obtained. 
With the help of a suitable open-closed string vertex
we define open-string gauge invariant operators parametrized
by on-shell closed string states. We use regulated vacuum
string field theory to sketch how pure closed string amplitudes
on surfaces without boundaries
arise as correlators of such gauge invariant operators.

\end{abstract}

\tableofcontents


\sectiono{Introduction and Summary}\label{s0}

Since it became clear that open string field theory (OSFT)
\cite{OSFT}
could provide striking evidence \cite{KS} for the tachyon conjectures
governing the decay of unstable D-branes or the annihilation
of D-brane anti-D-brane pairs \cite{conj},  the intriguing possibility
of formulating string field theory
directly around the tachyon vacuum has
attracted much attention.  A proposal for such vacuum string field
theory (VSFT) was made in \cite{0012251}, and investigated in
\cite{0102112,0105058,0105168,0106010,0105059,0106036,
0105129,0105184,0107101,0108150,0110124,0110136,0110204,
0111034,0111069,0111087}.

Lacking an exact analytic solution of OSFT that
describes the tachyon vacuum, it has not been possible yet
to confirm directly, or to derive the VSFT action
from first principles.
Therefore VSFT has
been
tested for consistency.  The main property of VSFT is that
the kinetic operator, which in OSFT is the BRST operator, is
chosen to be independent of the matter conformal theory, and
is thus built only using the reparametrization ghost conformal
field theory.  Families of consistent candidates for this kinetic
term, many of which are  related via field redefinitions,
were exhibited in \cite{0012251}. It was possible to show that in VSFT
the ratios of tensions of D-branes are correctly reproduced
from the classical solutions purporting to represent such D-branes.
This was seen in  numerical experiments \cite{0102112}, and
analytically using a boundary conformal field theory (BCFT) analysis
whose key ingredient was the
construction of the sliver
state \cite{0105168} associated with a general BCFT.
The sliver is  a projector in the star algebra of
open
strings; its matter part is identified with the
matter part of the solution representing a D-brane
\cite{0102112,0006240,0008252}. These tests did not select a particular
kinetic
term, in fact, data concerning the kinetic term, as long as it is only
ghost
dependent, cancel
in the computation of ratios of tensions.

The formulation of VSFT cannot be considered complete unless
a choice is made for its kinetic term.  This choice seems necessary
in order
to understand confidently issues related to: (a) the normalization
of the action giving us the brane tensions, (b) the spectrum of states
around classical solutions, and (c) the emergence of closed string
amplitudes.
It has been suggested by Gross and Taylor \cite{0106036} and
by Schnabl \cite{schnabl} that  
it may be difficult to obtain solutions of VSFT with non-zero action
if we insist on finite normalization of the kinetic term, leading to the
conclusion that VSFT could be a workable,
but singular limit of a better defined theory.
Even if this is the case,
it is important to find
which particular choice of the kinetic term appears in this limit, and
to investigate it thoroughly.

Indeed this is what we shall do in the present paper. We are led
by various pieces of evidence to a specific form of $\QQ$ that
is quite canonical and interesting.  We should say at the
outset that this form of $\QQ$ leads to vanishing action for
classical solutions unless its overall normalization is taken
to be infinite. Hence
regulation appears to be necessary, and
as we shall discuss, possible. $\QQ$ is
a ghost insertion at the open string midpoint.
More precisely it takes the form
\be
\label{qform}
\QQ  \propto
{1\over 2i} (c(i)-c(-i)) = c_0 -(c_2 + c_{-2}) + (c_4 + c_{-4})- \cdots
\end{equation}
The open string is viewed as the arc $|z|=1, ~\Im (z) >0$,
and thus
$z=i$ is the midpoint. The selected
$\QQ$ arises from a consideration of  the equations of motion in the
Siegel
gauge.
Again, there was early evidence, based on level expansion \cite{0012251},
that for a finite kinetic term
the Siegel gauge would yield zero action, and perhaps other gauges would
be more suitable. But in the spirit of the present paper, where we are
willing to
allow infinite normalization of the kinetic term,
the Siegel gauge is a good  starting point.  This
strategy was recently investigated in a stimulating paper by
Hata and Kawano
\cite{0108150}.
In the Siegel gauge the equation of
motion $\Psi + b_0 (\Psi *
\Psi)=0$
can be solved analytically not only in the matter sector, where
the matter sliver arises \cite{0008252,0102112}, but also in the ghost
sector\footnote{ The methodology was introduced
in
\cite{0008252},
but the correct expressions were given in \cite{0108150}.}. It is then
possible
to compute $\Psi * \Psi$ and determine $\QQ$ by requiring that
that $\Psi * \Psi$
takes the form $-\QQ\Psi$.
The authors of \cite{0108150}  obtained expressions that could be
analyzed
numerically
to glean the form of $\QQ$.  We have done this analysis
and obtained
evidence
that
the operator in \refb{qform} arises.

We can also do a rather complete
analytical study using BCFT techniques to obtain a solution of the
equations of motion with kinetic term given in \refb{qform}.
Here, as a first step we twist the ghost conformal
field theory stress tensor to obtain an auxiliary BCFT
where the ghost fields $(c,b)$ have spins
$(0,1)$.
This is clearly a natural operation in view of \refb{qform} since
local insertions of dimension zero fields are simple to deal with.
Moreover the resulting Virasoro operators commute with $b_0$ and the new
SL(2,R) vacuum coincides with the zero momentum tachyon. Analytic
treatment
of the string field equations of motion becomes possible by rewriting
the original equations in this twisted BCFT, and one finds that
the solution is simply the sliver of the twisted BCFT!
This geometrical approach gives
a directly calculable expression for the Neumann coefficients
characterizing the solution, as opposed to
the analytic solution  \cite{0108150} that
involves inverses and square roots of infinite
matrices.  We give
numerical evidence that the solutions are one and the same.

Given that the classical solution in OSFT describing the tachyon vacuum is
quite regular, one could wonder about the origin of the infinite
normalization factor that appears in the choice of our kinetic term. The
only way this could arise is if the variables of VSFT are related to
those of OSFT by a singular field redefinition. We give examples of
singular field redefinitions which could do this. They involve
reparmetrizations of
the open string coordinate which are symmetric about the mid-point and
hence preserve the $*$-product. We start with a $\QQ$ that is
sum of integrals of local operators made of matter and ghost fields, and
consider a  reparametrization  that has an
infinite squeezing factor
around the mid-point.
This transforms the various local
operators (if they are primary)
according to their scaling dimension, with the coefficient of
the lowest dimension operator growing at fastest rate. Thus if the
initial $\QQ$ contains a piece
involving the integral of $c$,
then under this  reparametrization the
coefficient of this operator
at the mid-point grows at the fastest
rate. This not only provides a mechanism for explaining how the
coefficient of the kinetic term could be infinite, but also explains how
a kinetic operator of the form $c(i)-c(-i)$ emerges under such field
redefinition even if the initial $\QQ$ contains combinations of matter and
ghost operators.
This scenario supports a viewpoint, stressed in \cite{0106036,dtalk},
that a purely ghost $\QQ$  is
a singular representative
of an equivalence class of kinetic
operators having regular representatives 
built from  matter and ghost operators. 
This singular limit is useful for some
computations, {\it e.g.} ratios of tensions of D-branes, but working
with a regular representative
may be necessary for other
computations like  the overall normalization of the tension.

While the
BRST operator $Q_B$ happens to be invariant under the action of the
reparametrization generators $K_n$ that
are symmetric about the
the string midpoint,
$\QQ$ is not invariant under an
arbitrary reparametrization of this type.
Nevertheless, being a midpoint insertion, it transforms naturally
under reparametrizations $z\to f(z)$
leaving invariant the midpoint.
$c(\pm i)$
simply scale with factors proportional to the inverse of the
derivatives of $f$ at $\pm i$.

The expression for $\QQ$ chosen here is rather
special in
that it is concentrated at the midpoint, and thus it would
seem to be an operator that cannot be treated easily by
splitting into left and right pieces.
In particular
the action of $\QQ$ on the identity string field is not well
defined.
One can define $\QQ$, however, as the limit of ghost
insertions $\QQ_\epsilon$
that approach symmetrically the midpoint as
$\epsilon\to 0$, so that $\QQ_\epsilon$
annihilates the identity for every non-zero $\epsilon$.
Although the action of
$\QQ_\epsilon$ on a state $|A\rangle$
can  be represented as $|S_\epsilon*A\rangle -
|A*S_\epsilon\rangle$ for an appropriate state
$|S_\epsilon\rangle$, and thus $\QQ_\epsilon$
would be seen to be an inner derivation,
the state $|S_\epsilon\rangle$, involving insertion of a
$c$ on the  identity string state just left of the midpoint, diverges as
$\epsilon$ approaches zero.
It thus seems unlikely that $\QQ$ can be viewed as an inner derivation.

\medskip
As mentioned above,
$\QQ$ defined this way has infinite normalization.
Via a field redefinition we could make $\QQ$ finite, at the cost of having
an infinite overall normalization of the VSFT action. In either
description regularization is necessary.
We examine this directly at the gauge fixed level. Working in the
Siegel gauge we introduce a parameter ``$a$" to define a deformation
$S(a)$ of the gauge fixed action, such that the VSFT action is
recovered for $a=\infty$. We
introduce a multiplicative factor
$\kappa_0(a)$ in front of the action to have a complete action $S_a
= \kappa_0(a) S(a)$.
In order to
have a succesful regularization we require
that $S(a)$ for any
fixed $a$ gives  a finite value for the energy of the classical solution
representing a D-brane. We find evidence that this is the case
using the level expansion procedure. The prefactor $\kappa_0(a)$ can then
be adjusted to give, by construction, the correct tension of the
D-brane solution.
The Feynman rules in this regulated VSFT generate correlation functions
on world-sheet with boundary,
with an additional factor involving a
boundary
perturbation, in close analogy with the effect of switching on constant
tachyon background in boundary string field theory (BSFT) \cite{bsft}.

The level expansion analysis of the VSFT equations of motion
leads to a surprise.
The numerical data indicates
that the solution is converging towards a projector that is different
from the sliver.
Like the sliver the new projector is a surface state.
Considering that the sliver represents only one of an infinite set of
projectors, this result is not totally unexpected.
We provide a list of a whole class of surface states, refered to as the
butterfly states,
satisfying properties similar to the sliver.
There is strong numerical evidence that
the solution in level expansion is approaching one particular member
of this class, $-$ a state which is a product of the vacuum state of the
left half string and the vacuum state of the right half-string.
Further properties of the butterfly states
are
currently under investigation \cite{wip}.
\medskip

Another subject we discuss in great detail is that of closed
strings.\footnote{For other attempts at getting
closed strings from open string theory around the tachyon vacuum, see
refs.\cite{closed,closedbsft,0111069}.}   
Our analysis begins with the introduction of gauge
invariant operators in OSFT. These open string field operators
$\OO_V(\Psi)$
are parametrized by on-shell closed string vertex operators $V$, and
concretely
arise from an open/closed transition vertex that emerged in studies
of closed string factorization in OSFT loop diagrams\cite{thorn}.
This open/closed vertex was studied geometrically in \cite{9202015}
where it was shown that supplemented with the cubic open string
vertex it would generate a cover of the moduli spaces of surfaces
involving open and closed string punctures.
In OSFT the correlation functions
of such
gauge invariant operators gives us the S-matrix elements of the
corresponding closed string vertex opertors computed
by integration over the moduli spaces of surfaces with boundaries.

We argue
that in VSFT gauge invariant operators take
an identical form, and confirm that our choice of $\QQ$ is consistent
with this.  We then sketch how the correlation function of $n$
gauge invariant operators in regulated VSFT could be
related to the closed string S-matrix
of the $n$ associated vertex operators arising by integration over
moduli spaces of surfaces {\it without boundaries}.  This means that
conventional pure closed string amplitudes could
emerge from  correlators of gauge invariant
operators in VSFT. In this analysis
we begin by noting  that at the level of string diagrams
$a=0$ gives us back the usual OSFT Feynman rules, whereas
as we take the regularization parameter $a$ to $\infty$
this corresponds to
selecting a region of the moduli space
where the length of the boundary
is going to zero. By a scaling transformation, and a factorization
analysis we find  that the amplitude reduces to one involving
the $n$ closed string vertex operators and an
additional
zero momentum closed string vertex operator of dimension $\le  0$.
We show in detail how a new minimal area problem
guarantees that the string diagrams for
these $n+1$ closed string vertex
operators
do generate a
cover of the relevant moduli space
of closed Riemann surfaces. This shows
how closed string
moduli arise from the original open string moduli.
If the only contribution to this $(n+1)$-point amplitude comes from the
term where the additional closed string vertex operator is the zero
momentum dilaton, then we get back the $n$-point closed string amplitude
of the external vertex operators.
In the picture that emerges, closed string
states are not introduced by hand -- bulk operators of the CFT
(necessary to even define the BCFT in question) are used to write
open string functionals that represent the closed string states.
Those are simply the gauge invariant operators of the theory.
Since the analysis, however, is sensitive to the regularization procedure,
complete understanding will require a better control of the
regularization procedure.

In the final section of this paper we discuss our results,
offer some perspectives, and suggest two possible alternative
directions for investigating vacuum string field theory.
In one of them we propose a pure ghost action that is completely
regular, but the string field has to be of ghost number zero.
In the other we suggest that many of the successess of VSFT
may be preserved if we introduced bulk matter stress tensor
dependence at the midpoing into the kinetic term.

\sectiono{The Proposal for a Ghost Kinetic Operator} \label{s1}

In this section we state our proposal for the purely
ghost kinetic operator $\QQ$ in VSFT and discuss the novel gauge structure
that emerges.
The kinetic operator $\QQ$ is a
local insertion of a ghost field with infinite coefficient.
We explain
how such kinetic term could arise from less singular choices via
reparametrizations that map much of the string to its midpoint.
With this choice of kinetic term, the gauge symmetry  is enlarged 
as compared
with that of usual open string field theory. 
Finally we explain the sense in which $\QQ$ is not an inner derivation,
but can be viewed as the limit element of a set of derivations that are
inner.

\subsection{Ghost kinetic operator and gauge structure}

The conjectured action for vacuum string field theory is given
by \cite{0106010}:
\be \label{eg1}
S = -\kappa_0 \Big[ {1\over 2} \langle \Psi, \QQ \Psi\rangle +
{1\over 3} \langle \Psi, \Psi * \Psi \rangle \Big]\, ,
\end{equation}
where $\kappa_0$ is an overall normalization constant,
$\QQ$ is an operator
made purely of the ghost world-sheet fields,
$|\Psi\ra$ is the string field represented by a
ghost number one state in the matter-ghost BCFT, and $\la A, B\ra\equiv \la
A|B\ra$ denotes the BPZ inner product of the states $|A\ra$ and $|B\ra$.
If $\QQ$ is nilpotent, is a derivation of the $*$-algebra, and
satisfies the hermiticity condition $\langle \QQ A|B\rangle = -(-1)^A
\langle
A| \QQ|B\rangle$, then
this action is
invariant under the
gauge transformation:
\be \label{egg1}
\delta |\Psi\rangle = \QQ|\Lambda\rangle + |\Psi*\Lambda\rangle -
|\Lambda*\Psi\rangle\, .
\end{equation}
Although the constant $\kappa_0$ can be absorbed into a rescaling of $\Psi$,
this changes the normalization of $\QQ$. We shall instead choose a
convenient normalization of $\QQ$ and keep the constant $\kappa_0$ in the
action as in eq.\refb{eg1}.

A class of kinetic operators $\QQ$ satisfying the
required constraints for gauge invariance was
constructed in
\cite{0012251}. They have the form:
\be \label{ecl}
\QQ=\sum_{n=0}^\infty
u_n \CC_{2n}\, ,
\end{equation}
where the $u_n$'s are constants, and,
\ben \label{eg3}
\CC_n &=& c_n + (-1)^n c_{-n}\,  \quad n\not= 0,\nonumber \\
\CC_0 &=& c_0 \,.
\end{eqnarray}

\medskip
We propose the following form of $\QQ$ as a consistent and
canonical choice
of  kinetic operator of VSFT:
\ben \label{ep1}
\QQ &=&{1\over 2i} (c(i) - \bar c(i)) = {1\over 2i} (c(i) - c(-i)) =
\sum_{n=0}^\infty
(-1)^n \CC_{2n}\, , \nonumber\\
&=& c_0 - (c_2 + c_{-2}) +  (c_4 + c_{-4}) - \cdots \,.
\end{eqnarray}
With this choice of $\QQ$,
the overall
normalization $\kappa_0$ will turn out to be infinite,
but we shall discuss a
specific method for
regularizing this infinity.
In writing the expression for $\QQ$
we are using the standard procedure of using the double cover of the open
string world-sheet, with anti-holomorphic fields in the upper half plane
being identified to the holomorphic fields in the lower half plane.

It is instructive to discuss in which sense the cohomology
of $\QQ$ so defined vanishes.  In fact, the equation
$\QQ |\Psi\rangle=0$ has no solutions if $ |\Psi\rangle$
is a Fock space state. This is clear since any state built from
finite linear combinations of monomials involving finite number
of oscillators must have bounded level, while
$\QQ$ involves oscillators of all levels, including therefore infinitely
many oscillators that do not annihilate the state $ |\Psi\rangle$.
Therefore, there is no standard Fock-space open string
cohomology simply because there are no $\QQ$ closed
states in the Fock space.  Suppose on the other
hand that a more general state $|\chi\rangle$ is annihilated by $\QQ$.
Then, given that $\QQ$ contains $c_0$ with unit coefficient,
we have that $|\chi\rangle = \{ \QQ, b_0\} |\chi\rangle= \QQ (b_0
|\chi\rangle )$.
Two things should be noted: $|\chi\rangle$ also has a local insertion
at the string midpoint, and, it appears to be always $\QQ$ trivial.
The only subtlety here is that $(b_0 |\chi\rangle)$ could be infinite in
which case the triviality of $|\chi\rangle$ is questionable. In fact, this
possibility arises for the case of gauge invariant operators related to
closed strings, as will be discussed in section \ref{s5}.

As discussed in earlier papers,
field redefinitions relate many of the kinetic terms of the form
\refb{ecl}. Typically these field redefinitions are induced by
world-sheet
reparametrization symmetries which are symmetric around the string
mid-point, and leaves the mid-point invariant. Such a reparametrization
$z\to f(z)$,
while acting on the kinetic operator of the form \refb{ep1}, will
transform $\QQ$ to
$(2i)^{-1}((f'(i))^{-1}c(i) - (f'(-i))^{-1}c(-i))$.
This
leaves $\QQ$ invariant if $f'(i)=1$. Thus we see that
the kinetic operator $\QQ$ is actually invariant under a
complex codimension 1
subgroup of the group generated by the $K_n$
transformations.

The choice of kinetic term is special enough that
the action defined by
\refb{eg1} and
\refb{ep1}, generally invariant only under the gauge
transformations in\refb{egg1}, is in fact invariant under two
separate sets of
gauge invariances:
\be \label{eg1aa}
\delta|\Psi\rangle = \QQ |\Lambda\rangle \, ,
\end{equation}
and
\be \label{eg1ab}
\delta|\Psi\rangle = |\Psi*\Lambda\rangle -
|\Lambda * \Psi\rangle\, .
\end{equation}
In fact, the quadratic and the cubic terms in the action are
separately invariant under each of these gauge transformations.
These follow from the usual
associativity of the $*$-product, nilpotence
of $\QQ$, and the
additional relation:
\be \label{eadd1}
\langle \QQ A, B * C\rangle = 0\, ,
\end{equation}
which holds generally for arbitrary Fock space states
$A,B$ and $C$ (other orderings, such as
$\langle A, \QQ B * C\rangle$ and $\langle A, B * \QQ C\rangle$
also vanish).  This
relation in turn follows from the fact that
$\QQ$ involves
operators $c,\bar c$ of dimension $-1$ inserted at $i$. As a result:
\be \label{add2}
\langle \QQ A, B * C\rangle = \langle f_1\circ (\QQ A(0)) f_2\circ B(0)
f_3\circ C(0)\rangle
\end{equation}
vanishes since
the conformal transformation of $\QQ$
gives a factor of $(f_1'(i))^{-1}$, and
$f_1'(i)$ is infinite.\footnote{For more
general states, such as surface states or squeezed states
the inner product \refb{add2}  might be nonvanishing. For
example, note that \refb{eadd1} {\it does not} imply
that $\langle A, \QQ B \rangle =0$ 
as might be suggested
by the identity $\langle A, \QQ B \rangle = \langle \II , A * \QQ
B\rangle$.
This latter expression does not vanish since the identity string
field
is not a Fock space state. We cannot therefore assume that
$A * \QQ B$ can be set to zero even for Fock space states $A$ and
$B$.} ($f_i$'s are the standard conformal maps appearing
in the definition
of the $*$-product and have been defined below
eq.\refb{es37}).
The symmetry of the action under the homogeneous transformation
\refb{eg1ab} is in accordance with the conjecture that at the tachyon
vacuum many broken symmetries should be
restored\cite{0009038,closedbsft}.

\subsection{Possible origin of a singular $\QQ$} \label{s2origin}

While the VSFT action described here is singular, the original OSFT action
is non-singular. In this subsection we shall attempt to understand how a
singular action of the type we have proposed could arise from OSFT.

In order to compare VSFT with OSFT, it is more convenient to
make a rescaling $|\Psi\rangle = (g_o^2 \kappa_0)^{-1/3} |\wt\Psi\rangle$ to
express
the action as:
\be \label{egx1}
S = - {1\over g_o^2} \Big[ {1\over 2} \langle \wt\Psi,
\wt\QQ
\wt\Psi\rangle +
{1\over 3} \langle \wt\Psi, \wt\Psi * \wt\Psi \rangle \Big]\, ,
\end{equation}
where
\be \label{egx2}
\wt \QQ = (g_o^2 \kappa_0)^{1/3} \QQ\, .
\end{equation}
Here $g_o$ is the open string coupling constant.
OSFT expanded around the tachyon vacuum solution $|\Phi_0\rangle$
has the same form except
that $\wt \QQ$ is replaced by the operator $\wh \QQ$:
\be \label{egx3}
\wh \QQ|A\rangle = Q_B|A\rangle + |\Phi_0*A\rangle -
(-1)^A |A*\Phi_0\rangle\, .
\end{equation}
Since $\kappa_0$ is infinite, so is $\wt\QQ$. On the other hand,
since the classical solution $|\Phi_0\rangle$ describing the tachyon
vacuum in OSFT
is
perfectly regular, we expect $\wh \QQ$ to be regular. Thus one could ask
how a singular $\wt\QQ$ of the kind we
are
proposing could arise. Clearly for this to happen the OSFT and the
VSFT
variables must be related by a singular field redefinition. We shall now
provide an example of such field redefinition which not only explains how
the coefficient of the kinetic term could be infinite, but also provides a
mechanism by which ghost kinetic operator proportional to $c(i)-c(-i)$
could arise.

Let us begin with a $\wt \QQ$ of the form:
\be \label{egx4}
\wt \QQ = \sum_r \int d\sigma a_r(\sigma) O_r(\sigma) \, ,
\end{equation}
where $a_r$ are smooth functions of $\sigma$ and $O_r$ are local operators
of ghost number 1, constructed from products of $b$, $c$, and matter
stress tensor. The above expression is written on
the double cover
of the strip so that $\sigma$ runs from $-\pi$ to $\pi$ and we only
have holomorphic fields.
Since $a_r$ are finite such a $\wt \QQ$
might
be obtained
from OSFT by a
non-singular
field redefinition. Given this $\wt \QQ$, we can generate other
equivalent $\wt \QQ$ by  
reparametrization of the open string coordinate $\sigma$ to $f(\sigma)$
such that $f(\pi-\sigma) = \pi - f(\sigma)$
for $0\le\sigma\le\pi$ and
$f(-\sigma-\pi)=-\pi-f(\sigma)$ for $-\pi\le\sigma\le 0$.
Such reparametrizations
do not change the structure of the cubic term but changes the kinetic
term.
If $O_r$ corresponds to a primary field of dimension $h_r$, then under
this reparametrization $\wt\QQ$ transforms to:
\be \label{eqnew}
\sum_r \int d\sigma a_r(\sigma) (f'(\sigma))^{h_r} O_r(f(\sigma)) \, .
\end{equation}
Consider now a reparametrization such that $f'(\pm\pi/2)$ is
small and in
particular $\int d\sigma (f'(\sigma))^{-1}$ gets a large
contribution from
the region around $\sigma=\pm\pi/2$.
Let us for example take $f'(\sigma)
\simeq (\sigma\mp{\pi\over 2})^2 + \epsilon^2$ for
$\sigma\simeq \pm
\pi/2$.\footnote{Another special case of this would be a choice of
$f(\sigma)$ where a finite region around the mid-point is squeezed to an
infinitesimal region.}
In this
case the dominant contribution to eq.\refb{eqnew} will come from the
lowest dimensional operator $c$
as long as the corresponding
$a_r$ does not vanish at $\pm \pi/2$,
and the transformed $\wt\QQ$ will be
proportional to
\be \label{eqnew2}
{1\over \epsilon} (c(\pi/2) +  c(-\pi/2) )\, .
\end{equation}
The relative coefficient between $c(\pi/2)$
and $c(-\pi/2)$ has been fixed by
requiring twist
invariance. In the upper half plane coordinates
($z=e^{\tau+i\sigma}$)
this  is proportional to
$\epsilon^{-1} (c(i) - c(-i))$, $-$ precisely the kinetic term of our
choice. Thus this analysis not only shows how a divergent coefficient
could appear in front of the kinetic term, but also explains how such
singular field redefinition could give rise to the pure ghost kinetic term
even if the original $\wt\QQ$ contained matter operators.

If $O_r$ is not a primary operator, then its transformation properties
under a reparametrization is more complicated. Nevertheless, given any
such operator containing a product of matter and ghost pieces, the
dominant contribution to its transform under a singular reparametrization
of the form described above will come from the lowest dimensional
operator, i.e. $c$ or $\bar c$, unless the coefficients of these terms
cancel between various pieces (which will happen, for example, if the
operator is a (total) Virasoro descendant of a primary other than $c$,
{\it e.g.} the BRST current).

To summarize, according to the above scenario the singular
$\wt\QQ$ of VSFT given in eq.\refb{egx2}, \refb{ep1} is
a singular member of an equivalence class of $\wt\QQ$'s whose generic member
is non-singular and is made of matter and ghost operators. The singular pure
ghost representative is useful for certain computations {\it e.g.} ratios of
tensions of D-branes, construction of multiple D-brane solutions etc.,
whereas the use of a regular member may be necessary for other computations
{\it e.g.} overall normalization of the D-brane tension, closed string
amplitudes etc.

Singular reparametrizations of the kind discussed above could also explain
the appearance of a sliver or other projectors as classical solutions of
VSFT. As an example we shall illustrate how an appropriate singular
reparametrization could take any finite $|m\rangle$ wedge state to the
sliver. From refs.\cite{0006240,0105168} we know that the wave-functional
of a wedge state $|m\rangle$ can be represented as a result of functional
integration on a wedge of angle $\alpha_m=2\pi(m-1)$ in a complex ($\wh
w$)  plane, bounded by the radial lines $\wh w = \rho e^{i{\pi\over 2}}$
and $\wh w = \rho e^{i({\pi\over 2} + 2\pi(m-1))}$. We put open string
boundary condition on the arc, and identify the lines $\wh w = \rho
e^{i{\pi\over 2}}$
and $\wh w = \rho e^{i({\pi\over 2} + 2\pi(m-1))}$ as the left and the
right halves of the string respectively. In particular if $\sigma$ denotes
the
coordinate on the open string with $0\le\sigma\le\pi$,  then the line $\wh
w = \rho
e^{i{\pi\over 2}}$ is parametrized as:
\be \label{eone}
\wh w = {1+i e^{i\sigma}\over 1 - i e^{i\sigma}} = i\tan({\pi\over 4} -
{\sigma\over 2})\, , \qquad \hbox{for} \quad 0 \le
\sigma\le{\pi\over 2}\, .
\end{equation}
{}From the above description it is clear that we can go from an wedge
state $|m\rangle$ to an wedge state $|n\rangle$ via a reparametrization:
\be \label{etwo}
(\wh w'/i) = (\wh w / i)^\gamma, \qquad \gamma = {\alpha_n\over \alpha_m}
= {n-1\over m-1}\, .
\end{equation}
In terms of $\sigma$, this corresponds to the transformation:
\be \label{ethree}
\tan({\pi\over 4} - {\sigma'\over 2}) = \tan^\gamma({\pi\over 4} -
{\sigma\over 2})\, .
\end{equation}
In order to get the sliver, we need to take the $n\to\infty$ limit. In
this limit $\gamma\to\infty$. Since for $0 < \sigma < \pi/2$, $|\tan(\pi/4
- \sigma/2)| < 1$, we see that as $\gamma\to\infty$ any $\sigma$ in the
range $0 < \sigma < \pi/2$ gets mapped to the point $\sigma'=\pi/2$. This
corresponds to squeezing the whole string into the mid-point. This shows
that using such singular reparametrization we can transform any wedge
state $|m\rangle$ into the sliver, and provides further evidence to the
conjecture that the VSFT action with pure ghost kinetic term arises from
OSFT expanded around the tachyon vacuum under such singular
reparametrization.

\subsection{Action of $\QQ$ on the identity state}

Among the constraints for gauge invariance is the derivation
property
\be
\label{der}
\QQ ( A* B) =  \QQ A * B + (-1)^A \, A * \QQ B
\end{equation}
which must hold. This property indeed holds for
each $\CC_n$ and therefore holds
for the chosen $\QQ$.
On the other hand, there was a criterion related to the
identity string field $\II$ that distinguished  two
classes of kinetic operators.
There are candidate operators for $\QQ$ which
viewed as integrals over the string, have vanishing support
at the string midpoint. Those operators annihilate the
identity and they split into left and right parts, as
discussed in \cite{0106036}. On the other hand there
are operators, such as $c_0$, which do not kill the
identity.\footnote{Since they are
derivations
of the star algebra we believe
they should be viewed as outer derivations. Indeed, an
inner derivation $D_\epsilon$ acts as $D_\epsilon A =
\epsilon * A - A * \epsilon$, so it is reasonable to demand
that all inner derivations annihilate the identity string field.
Not being inner, $c_0$ would have to be outer.}
As we will discuss below,
our choice of $\QQ$, using insertions precisely at the midpoint,
does not annihilate the identity.
In fact, a direct computation shows that
$\QQ \II $ is divergent.
However, we will also
show that $\QQ$ can be
considered as the limit of a sequence such that every member of the
sequence
annihilates the identity state.

Recall that
$|\II\rangle$ is defined through the relation:
\be \label{edefid}
\langle \II |\phi\rangle = \langle h_1\circ \phi(0) \rangle_{D}
\end{equation}
for any Fock space state $|\phi\rangle$. Here
$h\circ\phi(0)$ denotes the conformal transform of the vertex operator
$\phi(0)$ by the conformal map $h$,
$\langle ~\rangle_{D}$ denotes correlation
function on a unit disk, and
the conformal map $h_N$ for any $N$ is
defined as
\be \label{eg15}
h_N(z) = \Big( {1 + i z\over 1 - i z}\Big)^{2/N}\, .
\end{equation}
Thus
$\la\II|c(i)|\phi\ra=\la h_1\circ c(i) h_1\circ \phi(0)\ra$ is
divergent since $h_1\circ c(i)=h_1'(i)^{-1} c(0)$ and $h_1'(i)$ vanishes.

We now
define a new operator $\QQ_\epsilon$ by
making the replacements
\ben
\label{replace}
c(i) & \to & {1\over 2} \Bigl(e^{-i\epsilon} c(ie^{i\epsilon}) +
e^{i\epsilon}c(i e^{-i\epsilon}) \Bigr), \nonumber\\
c(-i) &\to & {1\over 2}
\Bigl(e^{-i\epsilon} c(-ie^{i\epsilon}) +
e^{i\epsilon}c(-i e^{-i\epsilon}) \Bigr),
\end{eqnarray}
in \refb{ep1}.
In the local coordinate picture
where the open string is represented by the arc $|\xi| = 1$ in the upper
$\xi$ half-plane, this corresponds to splitting the midpoint insertion
$c(\xi= i)$ into two
insertions, one on the left-half and the other on the right-half of
the string. The splitting is such that for $\epsilon\to 0$ we recover
the midpoint insertion, but this time
\be
\label{anni}
\langle\II | \Bigl(e^{-i\epsilon} c(ie^{i\epsilon}) +
e^{i\epsilon}c(i e^{-i\epsilon}) \Bigr)  = 0 \,,
\end{equation}
as can be verified using equations \refb{edefid}, \refb{eg15} -- the
point
being that by the geometry of the identity conformal map the two
operators
land on the same point but with opposite sign factors multiplying them,
and
thus cancel each other out exactly.

The replacements \refb{replace} in \refb{ep1} lead to 
the operator $\QQ_\epsilon$:
\ben \label{ep1a}
\QQ_\epsilon &=& {1\over 4i}
\Bigl(
e^{-i\epsilon}c(ie^{i\epsilon})+e^{i\epsilon}c(i
e^{-i\epsilon}) -e^{-i\epsilon}
c(-ie^{i\epsilon})-e^{i\epsilon}c(-ie^{-i\epsilon})\Bigr)
\nonumber \\ &=&
\sum_{n=0}^\infty
(-1)^n \CC_{2n}\cos(2n\epsilon)\, .
\end{eqnarray}
Because of \refb{anni}, and an analogous result for the split version
of $c(-i)$, the operator
$\QQ_\epsilon$ defined above annihilates the identity
$|\II\rangle$ for every $\epsilon\ne 0$.   In addition,
being a superposition
of the anti-commuting
derivations $\CC_n$, it squares to zero, and
is a derivation.
It also has the expected BPZ property
\be
\langle \QQ_\epsilon A , B \rangle = -(-)^A \langle A,
\QQ_\epsilon B\rangle\,.
\end{equation}
Therefore, $\QQ_\epsilon$ satisfies all the conditions
for gauge invariance.

Since
$\QQ_\epsilon$ approaches the
$\QQ$ defined in \refb{ep1} as $\epsilon\to 0$, we could  {\it define}
\be
\label{tdef}
\QQ \equiv \lim_{\epsilon\to 0} \QQ_\epsilon \,.
\end{equation}
Defined in this way, $\QQ$ annihilates the identity. Acting on
Fock space states, such care is not necessary, and we can simply
use \refb{ep1}, but in general the
definition \refb{tdef} is useful.

We can express the action of $\QQ_\epsilon$ on a state $|A\rangle$ as
an inner derivation:
\be \label{eneweq3}
\QQ_\epsilon|A\rangle = |S_\epsilon * A\rangle -
(-1)^A |A * S_\epsilon\rangle\, ,  
\end{equation}
where
\be \label{eneweq4}
|S_\epsilon\rangle = {1\over 4i}\Big(e^{-i\epsilon}c(ie^{i\epsilon}) -
e^{i\epsilon}c(-ie^{-i\epsilon}) \Big)|\II\rangle\, .
\end{equation}
However note that $|S_\epsilon\rangle$ diverges in the $\epsilon\to 0$
limit since $\langle S_\epsilon|\phi\rangle$ for any Fock space state
$|\phi\rangle$ involves $(h_1'(i e^{i\epsilon}))^{-1}$, which diverges as
$\epsilon\to 0$. Thus while the $\QQ_\epsilon$ operators may be viewed as
inner derivations
for $\epsilon\not= 0$,
it does not follow that $\QQ$ can also be viewed as an inner derivation.

\sectiono{Algebraic Analysis of the Classical Equations}
\label{s2}

In this section we reconsider the algebraic analysis 
of the classical equations of motion of VSFT in the Siegel
gauge carried out in refs.\cite{0108150,0008252}. The main result
of the analysis of \cite{0108150} is an expression for the coefficients
$u_n$ defining $\QQ$ (see \refb{ecl}) in terms of
infinite matrices of Neumann coefficients in the ghost sector
of the three string vertex.
We shall summarize briefly  their results and evaluate numerically
the coefficients $u_n$, finding striking evidence that the $\QQ$ that
emerges is indeed that in \refb{ep1}.

\bigskip
As usual we begin by looking for a factorized solution:
\be \label{es21}
\Psi = \Psi_g \otimes \Psi_m\,,
\end{equation}
with $\Psi_g$ and $\Psi_m$ denoting ghost
and matter pieces respectively,
and satisfying:
\be \label{es22}
|\Psi_m\rangle = |\Psi_m *^m \Psi_m\rangle\, ,
\end{equation}
and
\be \label{es23u}
\QQ|\Psi_g\rangle + |\Psi_g*^g \Psi_g\rangle = 0\, .
\end{equation}
If we start with a general class of kinetic operators of the form
\refb{ecl} with $u_0$ normalized to one, and make the Siegel
gauge choice
\be \label{eg8}
b_0 |\Psi\rangle = 0\, ,
\end{equation}
then the equation of motion \refb{es23u} takes the form
\be \label{es23}
|\Psi_g\rangle + b_0  |\Psi_g *\Psi_g\rangle = 0\, .
\end{equation}
Note that these contain only part of the equations \refb{es23u} which are
obtained by varying the action with respect to fields satisfying
the Siegel gauge condition. The full set of equations will be used later
for determining $\QQ$.

The solution for $|\Psi_m\rangle$ can be
taken to be any of the solutions
discussed in
\cite{0105058,0105168,0106010,0105059}.
The solution for
$|\Psi_g\rangle$ is given
as follows. Denote the ghost part of the 3-string vertex
as:\footnote{The coefficients $\wt V^{rs}_{nm}$ are related
the ghost Neumann functions $\wt N^{sr}_{mn}$ introduced in
ref.\cite{gj2} as $\wt  V^{rs}_{nm}=-n\, \wt N^{sr}_{mn}$.}
\be \label{es23a}
|V_g\rangle_{123} = \exp\Big(\sum_{r,s=1}^3\sum_{n\ge 1, m\ge 0}
\,
c_{-n}^{(r)}\,
\wt V^{rs}_{nm} b_{-m}^{(s)}\Big) \prod_{r=1}^3 (c^{(r)}_0 c^{(r)}_1)
|0\rangle_{(1)} \otimes |0\rangle_{(2)}\otimes |0\rangle_{(3)}\, ,
\end{equation}
where $c^{(r)}_n$, $b^{(r)}_n$ are the ghost oscillators associated with
the $r$-th string and $|0\rangle_{(r)}$ denotes the SL(2,R)
invariant ghost vacuum of the $r$-th string. The matrices $\wt
V^{rs}_{mn}$
have
cyclic
symmetry $\wt V^{rs}_{mn}=\wt V^{r+1,s+1}_{mn}$
as usual. We now
define:
\ben \label{es23b}
&& (\wt V_0)_{mn}=\wt V^{rr}_{mn}, \qquad (\wt V_\pm)_{mn}=\wt
V^{r,r\pm1}_{mn},
\qquad \wt C_{mn}=(-1)^m \delta_{mn},
\nonumber \\
&& (\wt v_0)_m = \wt V^{rr}_{m0}, \qquad (\wt v_\pm)_m = \wt V^{r,
r\pm 1}_{m0},
\qquad \hbox{for} \quad 1\le m,n < \infty\, , \nonumber \\
\end{eqnarray}
\be \label{es23d}
\wt M_0 = \wt C \wt V_0, \qquad \wt M_\pm = \wt C \wt V_\pm\, ,
\end{equation}
and,
\be \label{es23e}
\wt T = {1\over 2 \wt M_0} \Bigl( 1 + \wt M_0 - \sqrt{(1-\wt M_0) (1+ 3
\wt M_0)}\Bigr), \qquad \wt S = \wt C \wt T\, .
\end{equation}
The solution to eq.\refb{es23} is then given by:
\be \label{es24}
|\Psi_g\rangle = \NN_g \exp \Bigl(\sum_{n,m\ge 1} c_{-n} \wt S_{nm}
b_{-m}
\Bigr) \,
c_1|0\rangle\, ,
\end{equation}
for some appropriate normalization constant $\NN_g$.
Given the solution $|\Psi_g\rangle$, one can explicitly construct
$|\Psi_g*^g\Psi_g\rangle$. It was shown in ref.\cite{0108150} that
\be \label{es25}
|\Psi_g *^g \Psi_g\rangle = -\Big (c_0 + \sum_{n\ge 1} \,
u_n
\,
\CC_n \Big)
|\Psi_g\rangle,
\end{equation}
where
the vector $u=\{u_1,u_2,\ldots \}$ is given by:
\ben \label{es26}
u &=& (1-\wt T)^{-1} \Big[\,\,\, \wt v_0   \nonumber \\
&&\hskip-55pt + (\wt M_+ , \wt M_-)
(1-\wt M_0)^{-1} (1+\wt T)^{-1} 
\begin{pmatrix}
1-\wt T \wt M_0 & \wt T \wt M_+
\cr \wt T \wt M_- & 1 - \wt T \wt M_0
\end{pmatrix} 
\wt T
\begin{pmatrix}
\wt v_+ \cr \wt v_-
\end{pmatrix}
\, \Big]\,.
\end{eqnarray}
This expression was simplified in refs.\cite{0110124,0111034}, but we use 
eq.\refb{es26} for our analysis.  
Using eqs.\refb{es23u} and
\refb{es25} we see that $\QQ$ can be identified as:
\be \label{es27}
\QQ = c_0 + \sum_{n\ge 1}
u_n \CC_n\, .
\end{equation}

The coefficients $\wt V^{rs}_{mn}$ and hence the matrices $\wt M_0$,
$\wt M_\pm$,
$\wt T$ and the vectors $\wt v_0$, $\wt v_\pm$ can be calculated using
the
results of \cite{gj2}. In turn, this can be used to calculate $u_n$ from
eq.\refb{es26}. For odd $n$, $u_n$ vanishes
by twist symmetry.
The numerical results for $u_n$'s
for even $n$ at different
levels of
approximation, and the values extrapolated to
infinite level using a fit,
have been shown in table \ref{t1}. The results are clearly consistent
with
$u_{2 n}=(-1)^n$
and hence the choice of $\QQ$ given in
\refb{ep1}.

\begin{table}
\begin{center}\def\st{\vrule height 3ex width 0ex}
\begin{tabular}{|l|l|l|l|l|l|} \hline
$L$ & $f_2$ & $ f_4$ & $ f_6$  &
$f_8$ & $f_{10} $
\st\\[1ex]
\hline
\hline
40 & -0.87392 & 0.830099 & -0.803468 &  0.784561 & -0.770184
\st\\[1ex]
\hline
80 & -0.881488 & 0.840223 & -0.814839 &  0.796433 & -0.781999
\st\\[1ex]
\hline
160 & -0.888335 & 0.849592 & -0.825743 & 0.808389 & -0.7947
\st\\[1ex]
\hline
240 & -0.892017 & 0.85465
 & -0.831672 & 0.814956 & -0.801763
\st\\[1ex]
\hline
320 & -0.894496 & 0.858053& -0.835666 & 0.819388 & -0.806544
\st\\[1ex]
\hline
$\infty$ & -0.97748 & 0.96864 & -0.961296 &
0.953502  & -0.944372
\st\\[1ex]
\hline
\end{tabular}
\end{center}
\caption{ Numerical results for $f_{2n}$ at different level
approximation.
The last row
shows the interpolation of the various results
to $L=\infty$, obtained via a fitting function of the
form $a_0 + a_1/\ln(L) + a_2/(\ln(L))^2+
a_3/(\ln(L))^3 $.} \label{t1}
\end{table}

\sectiono{BCFT Analysis of Classical Equations of Motion} \label{s3}

In this section we shall discuss a method of solving the equations of
motion \refb{es23u} using the techniques of boundary conformal field
theory. As a first step we introduce a twisted version of the ghost
CFT where the ghost field $c(z)$ is of dimension zero. We also establish
a one to one map between the states of the twisted and untwisted
theory. We then study the star product in the twisted theory and relate
it to that in the untwisted theory. The upshot of this analysis is that
with our $\QQ$ the ghost part of the standard string field  equations
is solved by the state 
representing
the sliver of the twisted ghost CFT.  We conclude with a direct CFT
construction of the Fock space representation of this twisted sliver
and find that it compares well with the algebraic
construction of the solution \cite{0108150}.

\subsection{Twisted Ghost Conformal Theory}

We introduce a new conformal field theory CFT$'$ 
by changing the energy momentum tensor on the strip as
\be \label{es31}
T'(w) = T(w) - \p j_g(w), \qquad \bar T'(\bar w)
= \bar T(\bar w) - \bar
\p
\bar j_g(\bar w)\, ,
\end{equation}
where $T$, $\bar T$ denote the energy momentum tensor in the original
matter-ghost system, $T'$, $\bar T'$ denote the energy
momentum tensor of new theory,
and $j_g=cb$,
$\bar
j_g=\bar c\bar b$
are the ghost number currents in the original theory.
The ghost operators in the new theory,
labeled as
$\B$, $\C$, $\bB$,  and $\bC$, to avoid confusion, have spins (1,0),
(0,0),
(0,1) and (0,0) respectively, and satisfy the usual boundary condition
$\B=\bB$,
$\C=\bC$ on the real axis.
A few important facts about the $b',c'$ system are given below:
\begin{itemize}

\item
The first order system $(b',c')$ has a central
charge of minus two.

\item Since $c'$ has dimension zero, the SL(2,R) vacuum $|0'\rangle$
of this system, defined as usual in the complex plane, satisfies
\be
\label{vac}
c'_{n\geq 1} |0'\rangle = 0\,.
\end{equation}

\item The Virasoro operators from $T'$ commute with $b_0$.

\end{itemize}

\medskip
\noindent
The mode expansions of $T$, $T'$ and $j_g$ on the
cylinder
with coordinate $w=\tau+i\sigma$
(obtained from the double cover of the strip,
identifying the holomorphic fields at
$(\tau,\sigma)$ with anti-holomorphic fields at $(\tau, -\sigma)$ for
$-\pi\le\sigma\le 0$)
are given by:
\be \label{es32}
T(w)=\sum_n L_n
e^{-n w} - {c\over 24}, \quad T'(w)=\sum_n L_n' e^{-nw}-{c'\over 24},
\quad
j_g(w)=\sum_n j_n
e^{-nw},
\end{equation}
where $c=0$ is the total central charge of the original theory and
$c'=24$
is the total central charge of the auxiliary ghost-matter theory.
  It follows from \refb{es31} and \refb{es32} that
\be \label{es33}
L_n' = L_n + n j_n +\delta_{n,0}\, .
\end{equation}
In the path integral description the
euclidean  world-sheet actions
$\SSS$ and $\SSS'$ of the two   
theories are related as:
\be \label{ew1}
\SSS' = \SSS + {i\over 2\pi} \, \int
d^2 \xi \sqrt{\gamma} R^{(2)} (\vp+\bar\vp)\, ,
\end{equation}
where $\xi$ denotes the world-sheet coordinates, $\gamma$ denotes the
Euclidean world-sheet metric, $R^{(2)}$ is the scalar curvature computed
from the metric $\gamma$ and $\vp$, $\bar\vp$ are the bosonized ghost
fields
related
to the anti-commuting ghost fields as follows:
\be \label{ew2}
c \sim e^{i\vp}, \quad \bar c \sim e^{i\bar\vp}, \quad
b \sim e^{-i\vp}, \quad \bar b \sim e^{-i\bar\vp}\, ,
\end{equation}
and
\be \label{ew2p}
c' \sim e^{i\vp}, \quad \bar c' \sim e^{i\bar\vp}, \quad
b' \sim e^{-i\vp}, \quad \bar b' \sim e^{-i\bar\vp}\, .
\end{equation}
It should be noted that on general world-sheets the $\vp$ field has
different dynamics in the two theories.
On the strip, however, the world-sheet curvature vanishes and we have
$\SSS=\SSS'$.
The $\vp$ fields in the two theories can be identified,
and  hence the above equations allow
an identification of states in the two theories by the following map
between
the
oscillators and the vacuum states of the two theories:
\be \label{es34}
b_n \leftrightarrow b'_n, \qquad c_n \leftrightarrow c'_n, \qquad
c_1|0\rangle
\leftrightarrow |0'\rangle\,\qquad 
\langle 0| c_{-1}
\leftrightarrow \langle 0'|\,,
\end{equation}
where $|0\rangle$ and $|0'\rangle$ denote the SL(2,R) invariant vacua in
the original theory and the auxiliary theory respectively.
The
last two relations follows from the oscillator identification
and \refb{vac}.
We thus see that the zero momentum tachyon in the original
theory is the SL(2,R) vacuum of CFT$'$.
The two vacua
are normalized as
\be \label{es35}
\langle 0|c_{-1} c_0 c_1|0\rangle = \langle 0'|c'_0|0'\rangle =
V^{(26)} \, ,
\end{equation}
where $V^{(26)}$ denotes the volume of the 26-dimensional space-time. We
shall denote by $\langle \cdots \rangle$ and $\langle \cdots \rangle'$
the
expectation values of a set of operators in $|0\rangle$ and $|0'\rangle$
respectively. Also given a state $|A\rangle$ we shall denote by $A$ and
$A'$ the vertex operators of the state in the two theories in the upper
half plane coordinates. Thus:
\be \label{es36}
|A\rangle = A(0)|0\rangle=A'(0)|0'\rangle\, .
\end{equation}
Finally we note that BPZ conjugation in the twisted theory, obtained by the
map $|0'\ra\to\la 0'|$, $c_n \to (-1)^n c_{-n}$, and $b_n\to (-1)^{n+1}
b_{-n}$, can be shown to give a state identical to the one obtained by BPZ
conjugation in the original theory, given by $|0\ra\to \la 0|$, $c_n\to
(-1)^{n+1} c_{-n}$, and $b_n\to (-1)^n b_{-n}$. Thus we do not need to use
separate symbols for the BPZ inner product in the two theories.

\subsection{Relating Star Products and the analytic solution}

Next we would like to find the relationship between the star-products in
the two theories. We shall denote by $*$ and $*'$ the star products in
the
original and the auxiliary theory respectively. Thus:
\ben \label{es37}
\langle A| B * C\rangle &=& \langle f_1\circ A(0) f_2\circ B(0)
f_3\circ
C(0) \rangle, \nonumber \\
\langle A| B *' C\rangle &=& \langle f_1\circ A'(0)
f_2\circ B'(0) f_3\circ
C'(0) \rangle'\, ,
\een
where we have $f_1(z)=h_2^{-1}(h_3(z))$,
$f_2(z)=h_2^{-1}(e^{2\pi i/3}h_3(z))$,
and $f_3(z)=h_2^{-1}(e^{4\pi i/3} h_3(z))$, with $h_N(z)$ defined as in
eq.\refb{eg15}, are the standard conformal maps used in the
definition of
the
$*$ product.
The simplest way to relate these two star products is to use the path
integral prescription for $\langle A| B * C\rangle$ and $\langle A| B *'
C\rangle$ given in \cite{OSFT}. In
this description the star product is a
result of path integral over a two
dimensional surface in which three strips, each of of width $\pi$ and
infinitesimal length $\epsilon$, representing the external open strings,
are joined together such that the second half of the $r$-th string
coincides with the first half of the $(r+1)$-th string, for $1\le r\le
3$,
with the identification $r\equiv r+3$. The result is a flat world-sheet
with a single defect at the common mid-point of the three strings where
we
have a deficit angle of $-\pi$. Thus for this geometry $\int d^2\xi
\sqrt{\gamma}R^{(2)}$ gets a contribution of $-\pi$
from the mid-point,
and $\SSS$ and $\SSS'$ are
related as:
\be \label{es38}
\SSS' = \SSS - {i\over 2} (\vp(M) + \bar \vp(M))
\end{equation}
where $M$ denotes the location of the midpoint. Since the action appears
in the path integral through the combination $e^{-\SSS}$, we have
\be \label{es39}
\langle f_1\circ A(0) f_2\circ B(0) f_3\circ
C(0) \rangle = K_0 \langle f_1\circ A'(0) f_2\circ B'(0) f_3\circ
C'(0) \sigma^{+\prime}(M) \sigma^{-\prime}(M) \rangle'
\end{equation}
where $K_0$ is an overall finite normalization constant,
\be \label{epa3}
\sigma^{+\prime}=e^{i\vp/2}, \qquad \sigma^{-\prime} = e^{i\bar\vp/2},
\end{equation}
and $M=f_1(i)=f_2(i)=f_3(i)$. The primes
on $\sigma^{\pm\prime}$ denote
that these are operators in the auxiliary $b',c'$ system.
These operators
have conformal weights $(-1/8, 0)$ and $(0,-1/8)$ respectively.
We have explicitly verified eq.\refb{es39}
using specific choices of the states $|A\rangle$, $|B\rangle$,
$|C\rangle$. Since in the local coordinate system the mid-point of the
string is at $i$,
we can write  
\be \label{efour}
f_1\circ  A'(0) \sigma^{+\prime}(M) \sigma^{-\prime}(M)
= \lim_{\epsilon\to 0} |f_1'(i+\epsilon)|^{1/4} f_1\circ ( A'(0)
\sigma^{+\prime}(i+\epsilon)\sigma^{-\prime}(i+\epsilon) )\, .
\end{equation}
This, together with \refb{es37} and \refb{es39}, and the relation
$I\circ(\sigma^{+\prime}(i+\epsilon)\sigma^{-\prime}(i+\epsilon))
\simeq \sigma^{+\prime}(i-\epsilon) \sigma^{-\prime}(i-\epsilon)$ for 
$I(z)=-1/z$ gives
\ben \label{es310}
|B*C\rangle &=& \lim_{\epsilon\to 0} K_0 |f_1'(i+\epsilon)|^{1/4}
\sigma^{+\prime}(i-\epsilon) \sigma^{-\prime}(i-\epsilon) |B *' C\rangle
\,, \nonumber\\
&\propto &
\sigma^{+\prime}(i-\epsilon) \sigma^{-\prime}(i-\epsilon) |B *' C\rangle\,  ,
\end{eqnarray}
where the constant of proportionality is infinite since
$f_i'(i+\epsilon)\sim \epsilon^{-1/3}$ diverges as $\epsilon\to 0$.
However at this stage we are analyzing
the solution only up to a (possible infinite) scale factor, and so we
shall not worry about this normalization.

The equations of motion 
\be \label{eg7}
\QQ|\Psi\rangle +|\Psi*\Psi\rangle = 0\, ,
\end{equation}
can now be written as:
\be \label{eg311}
\QQ|\Psi\rangle \propto
- \sigma^{+\prime}(i-\epsilon)
\sigma^{-\prime}(i-\epsilon) |\Psi
*'\Psi\rangle \, .
\end{equation}
We shall show that for the choice of $\QQ$ given in \refb{ep1},
eq.\refb{eg311} is satisfied by a multiple of $\Xi'$ where $\Xi'$ is the
sliver of the auxiliary ghost - matter system, satisfying
\be \label{eg312}
\langle\Xi'| \phi\rangle = \langle f\circ
\phi'(0)\rangle'\, ,
\end{equation}
for any Fock space state $|\phi\ra$.
Here
$f(\xi)=\tan^{-1}\xi$. For this
we first need to know
what
form
the operator $\QQ$ takes in the auxiliary ghost-matter theory. We use
\ben \label{eg313}
\QQ &=& c_0 + \sum_{n\geq 1} (-1)^n (c_{2n}+c_{-2n})\nonumber \\
 &=&  c'_0 +\sum_{n\geq 1}
(-1)^n (\C_{2n}+\C_{-2n}) \nonumber\\
&=& \C(i)+\C(-i)
\, ,
\end{eqnarray}
where the argument of $c'(\pm i)$ in the
last two expressions refer to the
coordinates on the upper half plane. If we now take the inner product of
eq.\refb{eg311} with a Fock space state $\langle\phi|$, then for the
choice $|\Psi\rangle\propto |\Xi'\rangle$, the left hand side is
proportional to:
\be \label{eg314}
\langle f\circ\Big(\phi'(0) \big(\C(i)+\C(-i)\big)\Big)\rangle' =
\langle f\circ\phi'(0) \Big(\C(i\infty)+\C(-i\infty)\Big)\rangle' \, .
\end{equation}
Note that since $c$ has
dimension zero in the auxiliary BCFT,
there is no conformal factor in its transformation.
On the other hand, since $\Xi'*'\Xi'=\Xi'$,
and $f(i+\epsilon) \simeq -{i\over 2} \ln\epsilon \equiv i\eta$,
the right hand side is
proportional to
\ben \label{eg315}
\langle f\circ\Big(\phi'(0) \sigma^{+\prime}(i+\epsilon))
\sigma^{-\prime}(i+\epsilon)
\Big)\rangle' &\propto &
\langle f\circ\phi'(0)
\sigma^{+\prime}(i\eta)\sigma^{-\prime}(i\eta)\rangle' \nonumber\\
&\propto & 
\langle f\circ\phi'(0)
\sigma^{+\prime}(i\eta)\sigma^{+\prime}(-i\eta)
\rangle' \, ,\nonumber\\
\end{eqnarray}
where in the last expression
we have used the Neumann boundary condition on $\vp$ to relate
$\sigma^{-\prime}(i\eta)$ to $\sigma^{+\prime}(-i\eta)$.
Thus we need to show that \refb{eg314} and \refb{eg315} are equal up to
an overall normalization factor independent of $\langle\phi|$. This is
seen
as follows.
Since both correlators are being evaluated on the upper half plane, the
points $\pm i\infty$ correspond to the same points.\footnote{This can be
made manifest by making an SL(2,R) transformation that brings the point
at
$\infty$ to a finite point on the real axis.} Thus on the
right hand side
of eq.\refb{eg314} we can replace $\C(i\infty)+\C(-i\infty)$ by
$2\C(i\infty)$. On the other hand on the right hand side of \refb{eg315}
we can replace
$\sigma^{+\prime}(i\eta)\sigma^{+\prime}(-i\eta)$ by
the leading term in the operator product expansion of $\sigma^{+\prime}$
with $\sigma^{+\prime}$, i.e. $\C(i\infty)$.
As a result both \refb{eg314} and \refb{eg315} are proportional to
$\langle f\circ\phi'(0) \C(i\infty)\rangle'$.

At this point we would like to note that a different kind of 
star product has been analyzed in works
by Kishimoto \cite{0110124} and Okuyama \cite{0111087} which helps
in solving the Siegel gauge equations of motion in the oscillator 
formalism. It will be interesting to examine the relation between
the $*'$ product and the product discussed by these authors.

\subsection{The twisted sliver state from CFT and a comparison}

Since in the arguments above we have ignored various infinite
normalization factors, the result may seem  formal.
In this subsection, therefore, we verify
explicitly that the solution $\Xi'$ obtained this way agrees with the
solution obtained in refs.\cite{0008252,0108150} by algebraic method.
This has an added advantage. The geometrical construction of $\Xi'$
given below expresses the Neumann coefficients in terms of simple
contour integrals that can be evaluated exactly for arbitrary level.
On the other hand the algebraic solution, as usual, involves inverses
and square roots of infinite matrices, and therefore can only be
evaluated
approximately using level expansion.

This is done as follows. Writing $\Xi'=\Xi'_g\otimes \Xi_m$, we have for
the ghost part:
\be \label{eg317}
\langle\Xi'_g| = \wh\NN_g  \langle 0'|\exp\Bigl( -\sum_{n,m\ge 1} c_{n}
\bar S_{nm} b_{m}\Bigr)\,
\end{equation}
where $\wh\NN_g$ is a normalization factor. 
The BPZ dual ket is
\be \label{eg317b}
|\Xi'_g\rangle = \wh\NN_g  \exp \Bigl(\sum_{n,m\ge 1} c_{-n} \wh S_{nm}
b_{-m}\Bigr)\,
|0'\rangle, \qquad  \wh S_{nm} = (-1)^{n+m} \bar S_{nm}\,.
\end{equation}
The calculation of the matrix elements $\bar S$ (or $\wh S$) is done
using the a small variant of the CFT methods  in \cite{lpp}. 
The idea is to evaluate
\be
\label{eval2times}
h(z,w) \equiv \langle 0'|\exp\Bigl( -\sum_{n,m\ge 1} c_{n}
\bar S_{nm} b_{m}\Bigr)  c(w) b(z) \, c_0 |0'\rangle
\end{equation}
in two different ways. In the first one we expand using
$c(w) = \sum_p c_{-p} w^p$ and $b(z) = \sum_q b_{-q}z^{q-1}$ and find
\be
\label{eval1}
h(z, w) = -\sum_{n,m} w^m  z^{n-1} \bar S_{nm} \,\, \to \,\,
\bar S_{nm} = -\oint_0 {dz\over 2\pi i}{1\over z^n} \oint_0 {dw\over
2\pi i}
{1\over w^{m+1}} h(z,w) \,.
\end{equation}
In the second evaluation of \refb{eval2times} the right hand side is
viewed
as a correlator
\ben
\label{eval2}  
h(z,w) &=& \la f\circ c'(w) f\circ b'(z) f\circ c'(0) \ra' \nonumber \\
&=& \Bigl\langle  c'(f(w)) \, b'(f(z)) {df(z)\over dz} \, c'(f(0))
\Bigr\rangle' \nonumber \\
&=& {df(z)\over dz} \,{1\over f(w) - f(z)}\, {f(w) -f(0)\over f(z) -
f(0)}\,, 
\end{eqnarray}
where 
the function $f(\xi)$ denotes the insertion map associated with
the
geometry of the surface state, and the derivative ${df\over dz}$ arises
because
$b$ has conformal dimension one. The general result now follows from
comparison of
\refb{eval1} and \refb{eval2} together with \refb{eg317b}
\be \label{eg318a}
\wh S_{nm} =  (-1)^{n+m}\oint_0 {dz\over 2\pi i}{1\over z^n} \oint_0
{dw\over 2\pi
i} {1\over w^{m+1}} {df(z)\over dz} \,{1\over f(z) - f(w)}\, {f(w)
-f(0)\over
f(z) - f(0)} .
\end{equation}
This is the general expression for the Neumann coefficients of a
once punctured disk in the twisted ghost CFT. For our particular case,
the twisted sliver is defined by $f(z) = \tan^{-1} (z)$ and the Neumann
coefficients vanish unless $n+m$ is even. Therefore we find:
\be
\label{eg318}
\wh S_{nm} =  \ointop_0 {dz\over 2\pi i} {1\over z^n}\ointop_0
{d w \over 2\pi i}\,  {1\over w^{m+1}} {1\over 1+z^2} {1
\over(\tan^{-1}(z) -
\tan^{-1}(w))} {\tan^{-1}(w)\over \tan^{-1}(z)}\, .
\end{equation}
The first few coefficients are
\ben \label{ewhs}
&& \wh S_{11} = -\frac{1}{3} \cong -0.33333 \,,
\quad  \wh S_{31} =
\frac{4}{15} \cong
0.26667\,, \quad \wh S_{22} = \frac{1}{15} \cong 0.06667 \,,\nonumber\\
&& \wh S_{51} = -\frac{44}{189} \cong -0.23280 \,, \quad
\wh S_{33} = -\frac{83}{945} \,\cong -0.08783\,, \nonumber\\
&& \wh S_{42} = -\frac{64}{945} \cong -0.067724\,.
\end{eqnarray}
Since $|0'\rangle=c_1|0\rangle$, we see that
$|\Psi_g\rangle$ defined in eq.\refb{es24} and
$|\Xi_g'\rangle$
describe the same state if
the
matrices $\wt S_{mn}$ defined in eq.\refb{es23e}
and $\wh S_{mn}$ defined
in
eq.\refb{eg318} are the same.
The numerical results for $\wt S_{mn}$ are given in
table \ref{t2}, and can be seen to be in
good agreement with $\wh S_{mn}$
given in eq.\refb{ewhs}.

\begin{table}
\begin{center}\def\st{\vrule height 3ex width 0ex}
\begin{tabular}{ |l|  l|  l|  l|  l|  l| l |               } \hline
$L$ &   $\wt  S_{11}  $ &   $  \wt S_{31}  $   & $\wt S_{22}$  &
$\wt S_{51}  $ & $  \wt S_{33}  $    &   $\wt S_{42}$
\st  \\[1ex]
\hline
\hline
100 &    -0.31526  & 0.248339 & 0.066288 & -0.21432& -0.081680 &
-0.067263
\st\\[1ex]
\hline
150 & -0.316448 & 0.249482& 0.066269 &  -0.21543& -0.082030 & -0.067220
\st\\[1ex]
\hline
200 & -0.317244 & 0.250270 & 0.066271 &  -0.21621 & -0.082281 &
-0.067215
\st\\[1ex]
\hline
250 & -0.31783 & 0.250862& 0.066279 & -0.21680 & -0.082473 & -0.067220
\st\\[1ex]
\hline
$\infty$ & -0.33068 & 0.26345 & 0.067965 &
-0.22916   & -0.08642 & -0.06698
\st\\[1ex]
\hline
\end{tabular}
\end{center}
\caption{ Numerical results for $\wt S_{nm}$ at different level
approximation.
The last row
shows the interpolation of the various results
to $L=\infty$, obtained via a fitting function of the
form $a_0 + a_1/\ln(L)  $.} \label{t2}
\end{table}

\sectiono{Regularizing the VSFT action} \label{s4}

As pointed out already in section \ref{s1}, in order to get a
D-25-brane solution of finite
energy density, we need to take the overall
multiplicative
factor $\kappa_0$ appearing in eq.\refb{eg1} to be infinite. We shall now
discuss a precise way of regularizing the theory so that
for any fixed value of the regulator $a$, the value of $\kappa_0(a)$,
needed to
reproduce the D-25-brane tension correctly, is finite.
The action \refb{eg1}
is then recovered by taking the $a\to\infty$ limit.
Presumably this regularization
captures some of the physics of the correct regularization procedure
coming from the the use of  nearly singular reparametrization instead of
the
singular reparametrization discussed in section 
\ref{s2origin}.

\subsection{The proposal for regulated gauge fixed VSFT}

The regularization is done by first fixing the Siegel gauge
$b_0|\Psi\rangle=0$.  
In this way, the kinetic operator in
\refb{eg1}, with $\QQ$
given in \refb{ep1}, becomes $c_0$. We then replace this gauge
fixed kinetic
operator by $c_0(1 + a^{-1}L_0)$. The result is the regulated action
$S_a$
given by:
\be
\label{eg28cc}
S_a = -\kappa_0(a) \Big[ {1\over 2} \langle \Psi, c_0 (1 + a^{-1} L_0)
\Psi\rangle +
{1\over 3} \langle \Psi, \Psi * \Psi \rangle \Big]\, .
\end{equation}
The gauge fixed unregulated VSFT is recovered in the $a\to \infty$ limit. 

Although the parameter $a$ has been introduced as a
regulator, the
Feynman
rules derived from the regulated action \refb{eg28cc}
have some close resemblence to  
boundary string field theory (BSFT) rules \cite{bsft} 
in the presence of a constant tachyon background.
To see this, let us note that the propagator computed
from this action is proportional
to:
\be \label{es43}
{b_0\over L_0 + a} = b_0 \int_0^\infty dl e^{-l (L_0+a)} \, .
\end{equation}
This is similar to the propagator in OSFT except for the factor of
$e^{-la}$ in the integrand. The
three
string vertex computed from the action \refb{eg28cc}
is also proportional
to
the three string vertex of OSFT. Thus when we
compute the Feynman amplitudes using these
Feynman rules, we shall get an
expression similar to that in OSFT except for an additional factor of
$\exp(-a\sum_i l_i)$, where the sum over $i$ is taken over all the
propagators in the Feynman diagrams.
Now in OSFT a Feynman diagram can be
interpreted as correlation function on a Riemann surface obtained by
gluing strips of length $l_i$ using the three string overlap vertices.
Since each strip of length $l_i$ contributes $2l_i$ to the total length
of
the boundary in the Feynman diagram, we see that a factor of
$e^{-a\sum_i l_i}$ can also be interpreted as $e^{-a B/2}$, where $B$ is
the
total length of the boundary of the Riemann surface associated with the
Feynman diagram. This is reminiscent of the term $a\int d\theta$
representing
constant tachyon perturbation in the boundary string field theory, with
$\theta$ denoting the coordinate on the boundary. We should, however,
keep
in mind that the world-sheet metric used in defining constant tachyon
background in boundary SFT is different from the world-sheet metric that
appears naturally in the Feynman digrams of OSFT, and so we cannot
directly relate the
 tachyon of boundary SFT to the parameter $a$
appearing
here. Presumably the $a\to\infty$ limit corresponds to the same
configuration in both descriptions.

\subsection{Level truncation analysis}

To test the consistency of our regulation scheme,
we now perform a numerical analysis using the level
truncation approximation. We must find that for any fixed value
of the regulator $a$, computations
with the regulated
action (\ref{eg28cc}) have a well-defined finite limit
as the level of 
approximation $L$ is sent to infinity.
We define in the standard way the level
approximation $(L, 2L)$ by truncating
the string field up to level $L$ (level is
defined as $L \equiv L_0 + 1 $)
and keeping the terms in the action which have
a total level of $2 L$.
For a fixed value of $a$, and a given level of 
approximation $(L, 2L)$, we look for translationally invariant 
solutions  $\Psi^L_a$ in Siegel gauge 
corresponding to D-25 brane.

\begin{figure}[!ht]
\leavevmode
\begin{center}
\epsfysize=8cm
\epsfbox{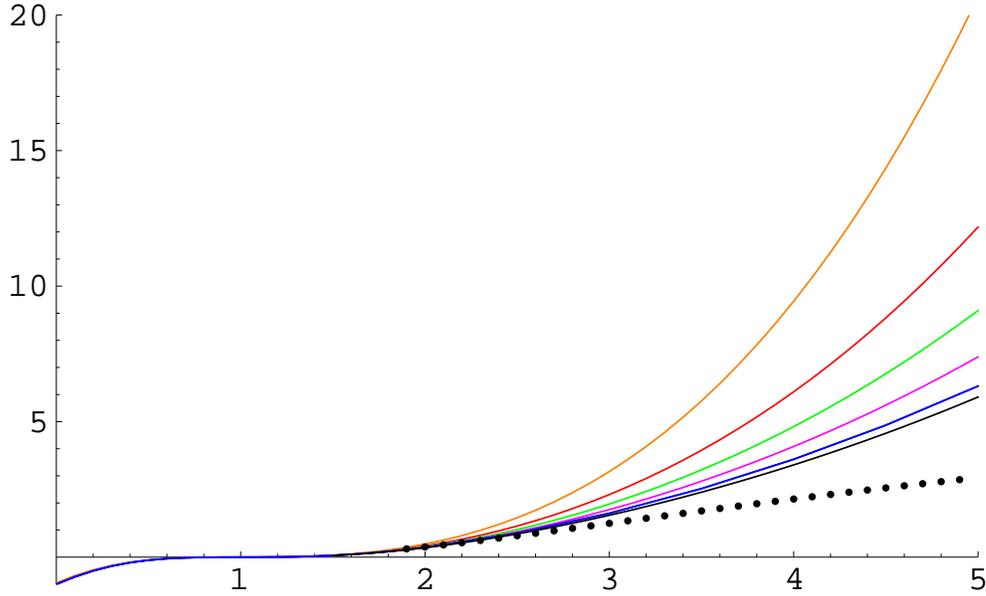}
\end{center}
\caption{This figure shows 
the plot of the function $a^3 f(a,L)$, 
computed at level $(L,2L)$ approximation, as a
function of $a$. Starting from the topmost graph, the six 
continuous curves correspond to
$L=2$,
4, 6, 8, 10 and 12 respectively. The lowermost
dotted curve is an $L =\infty$ extrapolation
of the data
obtained with a fit of the form $a_0 + a_1/L + a_2/L^2 + a_3/L^3$.
}
\label{f1} \end{figure}

The energy density of the D-25-brane solution in the
regulated theory at level $(L, 2L)$ 
approximation can be expressed as:
\be \label{es41}
\EE_a(L) = 
\frac{ \kappa_0(a)}{6} \la \Psi^L_a ,\,  (c_0 + a^{-1} L_0)
\Psi^L_a \ra \equiv \kappa_0(a) f(a,L)\, ,
\end{equation}
where $f(a,L)$ can be computed numerically.
It indeed turns out that for a fixed $a$, as the level of approximation
$L$ becomes larger than $a$, 
the function $f(a,L)$ approaches a finite value $F(a)$.
This is
best seen from Fig.\ref{f1}, where we have displayed the plot of $a^3
f(a,L)$
vs. $a$ for different levels of approximation $L$. 
Thus for a fixed
$a$, we get the energy density of the D-25-brane solution to be:
\be \label{es42}
\EE_a = \kappa_0(a) F(a)\, .
\end{equation}
We can
now take the $a \to \infty$
limit keeping $\kappa_0(a) F(a)$ to 
be fixed
at the D-25-brane tension $\TT_{25}$. In other words 
we choose the overall normalization of the action as 
\be
\kappa_0 (a) \equiv \frac{{\cal T}_{25}}{F(a)} \,.
\end{equation}
This gives a precise way of
defining the vacuum string field theory 
using level truncation scheme.

\medskip

If we go back to the analog of the $\wt\Psi$
variables by defining 
\begin{equation}
\Psi =
(\kappa_0(a) g_o^2)^{-1/3} \wt\Psi
\end{equation} 
then the action takes the form:
\be \label{ereg} 
S_a = - {1\over g_o^2} \, \Bigl[ {1\over 2}
\langle \wt\Psi, \, \wt \QQ_a \wt\Psi\rangle
+ {1\over 3} \langle \wt \Psi, \, \wt\Psi * \wt\Psi\rangle \Bigr]\, ,
\end{equation}
where
\ben \label{eqreg}
\wt\QQ_a &=& (g_o^2\kappa_0(a))^{1/3} (c_0 + a^{-1} c_0 L_0) 
=\bigg({g_o^2 \TT_{25}\over F(a)}\bigg)^{1/3} c_0 + \bigg({g_o^2
\TT_{25}\over
a^3
F(a)}\bigg)^{1/3} c_0 L_0 \, . \nonumber\\
\end{eqnarray}
The data in Fig.\ref{f1} 
suggests that  
$ a^3 F(a)$ grows linearly, {\it i.e.} 
$F(a) \sim 1/a^2$ for large $a$.
Hence, in the $a\to\infty$ limit,
the coefficient of  $c_0$ diverges, and that  of $c_0 L_0$ vanishes.
\begin{table}
\begin{center}\def\st{\vrule height 3ex width 0ex}
\begin{tabular}{|l|l|l|l|l|l|l|} \hline
$U(a,L)$ & $L=4$ & $L=6$ & $L=8$  &
$L=10$ & $L=12 $ & $L =\infty$
\st\\[1ex]
\hline
\hline
$a=2.0$  & -.0812 & -.0834 & -.0846 & -.0855 & -.0861 & -0.0904
\st\\[1ex]
\hline
$a=2.5$ & -.1121 & -.1164 & -.1191 &  -.1210 & -.1224 & -0.1315
\st\\[1ex]
\hline
$a=3.0$ & -.1372 & -.1436 & -.1478 & -.1508 & -.1530 & -0.1673
\st\\[1ex]
\hline
$a=3.5$ & -.1576 & -.1660 & -.1715 & -.1754 & -.1785 & -0.1994
\st\\[1ex]
\hline
$a=4.0$ & -.1744 & -.1845 & -.1911 & -.1959  & -.1996 &-0.2251
\st\\[1ex]
\hline
$a=4.5$ & -.1884 & -.1999 & -.2075 & -.2130 & -.2173 & -0.2468
\st\\[1ex]
\hline
$a=5.0$ & -.2002 & -.2130 & -.2214 & -.2275  & -.2323 &-0.2656
\st\\[1ex]
\hline
\end{tabular}
\end{center}
\caption{
Sample numerical results for the coefficient $U(a,L)$
at different level
approximation $(L,2L)$ for different values of $a$. 
}
\label{tn1}
\end{table}
\begin{table}
\begin{center}\def\st{\vrule height 3ex width 0ex}
\begin{tabular}{|l|l|l|l|l|l|l|} \hline
$V(a,L)$ & $L=4$ & $L=6$ & $L=8$  &
$L=10$ & $L=12 $ & $ L = \infty$
\st\\[1ex]
\hline
\hline
$a=2.0$  & -.1440 & -.1440 & -.1438 & -.1436 & -.1435 & -0.1429
\st\\[1ex]
\hline
$a=2.5$ & -.1884 & -.1887 & -.1886 &  -.1885 & -.1884 &-0.1878
\st\\[1ex]
\hline
$a=3.0$ & -.2225 & -.2232 & -.2234 & -.2234 & -.2234 &-0.2227
\st\\[1ex]
\hline
$a=3.5$ & -.2495 & -.2506 & -.2510 & -.2512 & -.2513 &-0.2515
\st\\[1ex]
\hline
$a=4.0$ & -.2712 & -.2728 & -.2735 & -.2738  & -.2740 &-0.2742
\st\\[1ex]
\hline
$a=4.5$ & -.2892 & -.2912 & -.2920 & -.2925 & -.2928 &-0.2946
\st\\[1ex]
\hline
$a=5.0$ & -.3043 & -.3066 & -.3076 & -.3082  & -.3086 &-0.3109
\st\\[1ex]
\hline
\end{tabular}
\end{center}
\caption{ Sample numerical results for the coefficient $V(a,L)$
at different level
approximation $(L,2L)$ for different values of $a$.
The last column of tables \ref{tn1} and \ref{tn2}
shows a large $L$ extrapolation obtained
with a fit $c_0 + c_1/L + c_2/L^2 + c_3/L^3$. The further
large $a$ extrapolation in (\ref{doublelimit}) 
is done with a more complete set of data than
shown in these tables (all values
of $a$ from 2 to 5 with an increment of 0.1). 
} 
\label{tn2}
\end{table}

We now examine the form of the D-25 brane solution.
The solutions $\Psi_L^a$ are string fields
belonging to the universal ghost number one
subspace \cite{conj,0006240}
obtained by acting on the vacuum
with ghost oscillators and matter Virasoro generators.
Up to level 4,
\ben
\Psi_a^L  = & T(a,L)  \Big[ & c_1 | 0 \ra + U(a,L) L^{m}_{-2} c_1 | 0 \rangle
+ V(a,L) c_{-1}|0 \ra   
 \\
&& \hskip-30pt  + A(a,L) L^m_{-4}c_1 | 0\ra   +  B(a,L)  
(L^m_{-2})^2 c_1 | 0 \ra + C(a,L) L_{-2}^m c_{-1} | 0 \ra + \nonumber\\
&&  \hskip-30pt D(a,L)( -3 \, c_{-3} + b_{-3}c_{-1} c_1) | 0\ra
  + E(a, L) b_{-2} c_{-2} c_1| 0 \ra 
+ \ldots 
\Big] \, .\nonumber
\end{eqnarray}
Our regulation prescription instructs
us to first take the large $L$ limit of $\Psi_a^L$,
and then remove the regulator by sending $a \to \infty$.
As shown in tables \ref{tn1} and \ref{tn2},
up to the overall normalization $T(a,L)$
which has been factored
out, the coefficients of the solution for a given
regulator $a$ are fairly stable
as the level is increased. Considering data
for $2 \leq a \leq 5$, and $L = 2, 4, 6, 8, 10, 12$,
we first perform a large $L$ extrapolation with a fitting
function of the form $c_0 + c_1/L + c_2/L^2 + c_3/L^3$; then
we extrapolate to $a=\infty$ with a fit $\gamma_0 +
\gamma_1/a + \gamma_2/a^2 + \gamma_3/a^3$.
This procedure gives
\ben \label{doublelimit}
\lim_{a \to \infty} \lim_{L \to \infty} 
\frac{ \Psi_a^L}{T(a,L)}  
& \cong &c_1 | 0 \ra  - 0.4564  \, L^{m}_{-2} c_1 | 0 \rangle
- 0.4901   
\, c_{-1} |  0 \rangle  \\  
&&\hskip-25pt  + \,  0.0041\, L^m_{-4} c_1 |0 \rangle +
0.0917\, (L^m_{-2})^2 c_1 | 0 \rangle + 0.2037 
\,L^m_{-2} c_{-1} | 0 \ra \nonumber
\\ && \hskip-25pt  -0.1131 (- 3 c_{-3}+ b_{-3} c_{-1} )|0\rangle   
- 0.0024 \,b_{-2}c_{-2} c_1| 0\ra + \ldots \nonumber
\end{eqnarray}
While this double extrapolation procedure is 
the correct general prescription,
we would like to show that
for certain purposes it is possible
to work in the non-regulated theory, or
in other words 
to commute the limits in (\ref{doublelimit}) 
by first removing the regulator sending $a \to \infty$
and then 
performing level truncation in the
theory with ${\cal Q} = (c(i) - c(-i))/(2i)$.
In fact, we know that the non-regulated
theory gives correct results about
existence of classical D-$p$ brane solutions
and the {\it ratios} of their tensions 
\cite{0102112,0105168} so it should
be the case that the limits in (\ref{doublelimit})
can be commuted for this class of physical questions. 
This
will obviously be the case if we can show
that {\it up to an overall normalization},
classical solutions are the same 
regardless of the order of limits,
\be \label{exchange}
\lim_{a \to \infty} \lim_{L \to \infty}
  \frac{\Psi_a^L}{T(a,L)} = \lim_{L \to \infty} \lim_{a \to \infty}
\frac{\Psi_a^L}{T(a,L)} \,.
\end{equation}
It is easy to perform numerical analysis directly at $a = \infty$
for a given level of approximation.
Although the energy, being proportional to $F(a)$, goes to zero in this
limit unless we compensate
for it by making $\kappa_0(a)$ large, the solution
approaches a finite limit up to the overall
normalization. Numerical results are shown in table \ref{tableainf}.

We find:
\ben \label{enum2}
\lim_{a \to \infty} \lim_{L \to \infty} 
\frac{ \Psi_a^L}{T(a,L)}  
& \cong &c_1 | 0 \ra  - 0.4603  \, L^{m}_{-2} c_1 | 0 \rangle
- 0.4900  
\, c_{-1} |  0 \rangle  \\  
&& \hskip-35pt + \,  0.0029\, L^m_{-4} c_1 |0 \rangle +
0.1049\, (L^m_{-2})^2 c_1 | 0 \rangle + 0.2311 
\,L^m_{-2} c_{-1} | 0 \ra \nonumber
\\ &&\hskip-35pt  -0.1258 (- 3 c_{-3}+ b_{-3} c_{-1} )|0\rangle   
+ 0.0001 \,b_{-2}c_{-2} c_1| 0\ra + \ldots \nonumber
\end{eqnarray}
This 
is compatible with (\ref{doublelimit}) 
and \refb{exchange}.

We thus find evidence that classical solutions of VSFT
are independent of the order of limits,
up to an overall normalization factor that needs to be adjusted
so as to keep the tension fixed.  This justifies
the analytic treatment of the equations
of motion based on matter/ghost factorization,
which has been an important assumption in
all studies of VSFT, and
which holds only in the $a \to \infty$ limit.
Moreover, we can study 
numerically the D-25 brane solution in the $a\to\infty$
theory at fixed $L$, which is much
simpler than taking the $L\to\infty$ limit
first and then taking the $a\to\infty$ limit.

\begin{table}
\begin{center}\def\st{\vrule height 3ex width 0ex}
\begin{tabular}{|l|l|l|l|l|l|l|l|} \hline
$$  &    $ U( L) $ & $ V( L)$ & $ A(L) $  &
$B( L) $ & $ C(L) $  & $ D(L) $ & $ E(L) $ 
\st\\[1ex]
\hline
\hline
$L=2$  & -.2879 & -.4576 &$\;$ -- & $\;$ -- & $\;$ --  &  $\;$ --  &  $\;$ --
\st\\[1ex]
\hline
$L=4$ & -.3015 & -.4357 & .0094 & .0358  & .1082  & -.0844 & -.0103
\st\\[1ex]
\hline
$L=6$ & -.3394 & -.4596 & .0080 & .0523 & .1440 & -.0995  & -.0037
\st\\[1ex]
\hline
$L=8$ & -.3631 & -.4708 & .0072 & .0627 & .1640 &  -.1072 & -.0019
\st\\[1ex]
\hline
$L=10$ & -.3798 & -.4771 & .0066 & .0700 & .1768& -.1114  & -.0011
\st\\[1ex]
\hline
$L=12$ & -.3923 & -.4811 & .0060 & .0755 & .1858 & -.1141  & -.0007
\st\\[1ex]
\hline
$L=\infty$ & -.4603 & -.4900 & .0029 & .1049 & .2311 &  -.1258 & .0001 
\st\\[1ex]
\hline
\end{tabular}
\end{center}
\caption{Coefficients of the $a = \infty$ solution,
at different level approximation $(L, 2L)$ (we use $U(\infty, L) \equiv U(L)$,
and he same convention for the other coefficients).
 The last row shows an extrapolation to infinite level
with a fitting function of the form $a_0 + a_1/L + a_2/L^2 + a_3/L^3$.
}
\label{tableainf}     
\end{table}

\begin{table}
\begin{center}\def\st{\vrule height 3ex width 0ex}
\begin{tabular}{|l|l|l|l|l|l|} \hline
$$  &    $ v_2 $ & $ \widetilde v_2$ & $ v_4 $  &
$\widetilde v_4 $ & $ v_6 $  
\st\\[1ex]
\hline
\hline
$L=2$  & -.2879 & -.4576 &$\;$ -- & $\;$ -- & $\;$ --
\st\\[1ex]
\hline
$L=4$ & -.3364 & -.4736 & .0056 & -.00193 & $\;$ --
\st\\[1ex]
\hline
$L=6$ & -.3655 & -.4816 & .0048 & -.00216 & -.00111
\st\\[1ex]
\hline
$L=8$ & -.3852 & -.4861 & .0043 & -.00197 & -.00080
\st\\[1ex]
\hline
$L=10$ & -.3999 & -.4891 & .0039 & -.00176 & -.00065
\st\\[1ex]
\hline
$L=12$ & -.4105 & -.4912 & .0036 & -.00157 & -.00056
\st\\[1ex]
\hline
$L=\infty$ & -.4778 & -.5027 & .0012 & -.00007 & -.0002
\st\\[1ex]
\hline
\end{tabular}
\end{center}
\caption{Coefficients of the $a = \infty$ solution written
as an exponential of matter and twisted ghost 
Virasoro operators, at different level approximation $(L, 3L)$.
 The last row shows an extrapolation to infinite level
with a fitting function of the form $a_0 + a_1/L + a_2/L^2 + a_3/L^3$.
}
\label{tablev}
\end{table}

It is illuminating to write the D-25 brane solution
in a basis of Fock states obtained by
acting on the zero-momentum tachyon 
with the matter Virasoro generators $L^m_{-n}$ 
and the ghost Virasoro generators 
$L^{' g}_{-n}$ ($n \geq 2$) 
of the twisted 
$bc$ system introduced in section \ref{s3}.\footnote{
A simple counting argument along the lines
of section 2.2 of \cite{0006240}
shows that 
all ghost number one Siegel gauge
string fields that belong to the $SU(1,1)$
singlet subspace \cite{0010190}
can be written in this form.} It turns out that to a very good
degree of accuracy the solution
can be written as 
\be
\Psi_{a =\infty} \sim \exp ( \sum_{n=1}^\infty v_{2n}L^m_{-2n})
 \exp ( \sum_{n=1}^\infty \widetilde v_{2n}L^{'g}_{-2n}) c_1 | 0 \ra .
\end{equation}
This is precisely the form expected for a 
surface state of the twisted BCFT  
introduced in section \ref{s3}.
The results for the coefficients $v_{2n}$ and $\widetilde v_{2n}$ at
various level approximations $(L, 3 L)$ are shown in
table \ref{tablev}. Extrapolating for $L \to \infty$
with a fit of the form $a_0 + a_1/L + a_2/L^2 + a_3/L^3$
we find
\ben \label{eresult}
\Psi_{a =\infty} & \sim & \exp( -0.5027 L^{' g}_{-2}
-0.00007 L^{'g}_{-4} + \dots ) c_1|0\rangle_g \nonumber \\
&& \otimes \exp(-0.4778 L^m_{-2}+
0.0012 L^m_{-4} -0.0002 L^m_{-6} + \dots
) |0\rangle_m\, .\nonumber\\
\end{eqnarray}
We note that although the
solution has precisely the form expected
for a surface state of the auxiliary matter-ghost system, 
it does not approach the twisted sliver $\Xi'$, 
for which the coefficient of $L'_{-2}$
is $-1/3$. This  should not bother us, however, since we
can generate many other surface states
(related to the sliver by a
singular or non-singular reparametrization of the string coordinate
symmetric about the mid-point)
which are all projectors. Moreover, at least formally,
all rank one projectors
are gauge-related in VSFT. 
The numerical result (\ref{eresult})
strongly suggests that 
as $L \to \infty$
the solution is in fact approaching
the remarkably simple state
\be
|B' \ra \sim \exp ( -\frac{1}{2} (L^m_{-2} + L^{'g}_{-2} ) c_1 | 0 \ra ,
\end{equation}
which we call the (twisted) {\it butterfly} state. 
It is possible to show that the 
state $|B' \ra$
is indeed a projector of the $*'$ algebra and 
an exact solution of the VSFT equations. In the
next section we shall come back to this point.

Let us finally check numerically that 
the Siegel gauge D-25 brane solution obtained in level truncation 
solves the equation of motion
of VSFT with our proposed $\QQ$. 
To this end we take the solution $\Psi_{a = \infty}$  
compute $\Psi_{a =\infty} * \Psi_{a = \infty}$, 
and try to determine $\QQ=(c_0 + \sum_{n\ge
1} u_{2n} (c_{2n} + c_{-2n}))$ up to a constant
of proportionality using the equation:
\be
\Psi_{a= \infty} * \Psi_{a =\infty}  \propto \QQ \Psi_{a =\infty} \, .
\end{equation}
The results for the coefficients $u_{2n}$ at various
level approximation $(L, 3L)$ are shown
in table \ref{tableu}
and are indeed consistent with our choice \refb{ep1} for $\QQ$.

 \begin{table}
\begin{center}\def\st{\vrule height 3ex width 0ex}
\begin{tabular}{|l|l|l|l|l|l|} \hline
$$  &    $ u_2 $ & $ u_4$ & $ u_6 $  &
$ u_8 $ & $ u_{10} $  
\st\\[1ex]
\hline
\hline
$L=2$  & -.8020 & $\;$--  &   $\;$--  &  $\;$--   &  $\;$-- 
\st\\[1ex]
\hline
$L=4$ & -.8672 & .7249 &  $\;$--   &  $\;$--  & $\;$ --
\st\\[1ex]
\hline
$L=6$ & -.9003 & .7918 & -.6854 &  $\;$--  &  $\;$-- 
\st\\[1ex]
\hline
$L=8$ & -.9201 & .8333 & -.7451  & .6615 &  $\;$--
\st\\[1ex]
\hline
$L=10$ & -.9334 & .8627 & -.7868 & .7138 & -.6457
\st\\[1ex]
\hline
$L=\infty$ &-.9969  & .9983 & -.923 & $\;$--   & $\;$-- 
\st\\[1ex]
\hline
\end{tabular}
\end{center}
\caption{Coefficients of the BRST operator
deduced from the $a = \infty$ solution, at
different level approximations $(L, 3L)$.The last row
shows an extrapolation to infinite level with fits
of the form $a_0 + a_1/L + a_2/L^2 + a_3/L^3$
($a_3 \equiv 0$ for $u_4$, $a_3 =a_2 \equiv 0$ for $u_6$).
}
\label{tableu}
\end{table}

\sectiono{The Butterfly State}

The level truncation results
have led to the discovery of a new
simple projector, the butterfly state,
different from the sliver.
There are
in fact several surface states that can be written
in closed form and shown to be projectors
using a variety of analytic approaches.
In this section we briefly state without proof some
of the relevant results. A thorough discussion will appear
in a separate publication \cite{wip}.

Consider the class of surface states
$|B_\alpha\rangle$, defined through:
\be \label{exy1}
\langle
B_\alpha|\phi\rangle \equiv \langle f_\alpha\circ \phi(0)\rangle_{D}
\end{equation}
with
\be \label{exy2}
f_\alpha(\xi) = {1\over \alpha} \sin(\alpha\tan^{-1}\xi) \, .
\end{equation}
As $\alpha \to 0$, we recover the sliver. For $\alpha = 1$
we have the butterfly state $|B \ra \equiv |B_{\alpha =1}\ra$,
defined by the map
\be
f_1 (\xi) = \frac{\xi}{\sqrt{1 + \xi^2}} \,.
\end{equation}
In
operator form the butterfly
can be written as
\be
| B\rangle =  \exp(-\frac{1}{2} L_{-2}) | 0 \ra \,.
\end{equation}
For any $\alpha$, these states can be shown to be 
idempotents
of the $*$ algebra,
\be
|B_{\alpha} \ra * |B_{\alpha} \ra =  |B_{\alpha} \ra \, .
\end{equation}
Moreover, in analogy with the sliver, the wave-functional
of $|B_\alpha\rangle$ factorizes into a product of
a functional of the left-half of the string and another functional of the
right half of the string. These states are thus naturally
thought as rank-one projectors in the half-string formalism
\cite{0105058, 0105059, 0106036}.
The key property that ensures
factorization is the singularity of the conformal
maps at the string midpoint,
\be \label{efpm}
f_{\alpha} (\pm i) = \pm i \infty \,.  
\end{equation}
It is possible to give a general argument
\cite{wip} that all sufficiently well-behaved
conformal maps with this property give
rise to split wave-functionals.

The case $\alpha =1$ is special  
because the wave-functional of the butterfly
$|B_{\alpha =1}\ra$ factors into the
product of the {\it vacuum} wave-functional of the right half-string and the
{\it vacuum} wave-functional of the left half-string. It is
thus in a sense the simplest possible projector. It is
quite remarkable that the same state
emerges in VSFT as the numerical solution preferred by
the level truncation  scheme.

Finally, in 
complete analogy
with the `twisted' sliver $\Xi'$, the `twisted'
states $|B'_{\alpha} \ra $
solve
the VSFT equations of motion with $\QQ = (c(i) - c(-i))/(2i)$,
\be 
\QQ |B'_{\alpha}\ra \propto  |B'_{\alpha} \ra *   |B'_{\alpha} \ra\,.
\end{equation}
Indeed the proof of section \ref{s3} 
that $\Xi'$ satisfies the VSFT
equations of motion $\QQ\Xi'\propto\Xi'*\Xi'$ only depends on the fact
that the map $f(\xi)=\tan^{-1}\xi$ associated with the sliver takes the
points $\pm i$ to $\pm i\infty$. As can be seen from \refb{efpm}, this
property is shared by the map $f_\alpha$ associated with the state
$|B_\alpha\ra$.


\sectiono{Gauge invariant operators in OSFT and VSFT} \label{s5}

Since open string field theory on an unstable D-brane has no physical
excitations at the tachyon vacuum, the only possible observables in this
theory are correlation functions of gauge invariant operators. A natural
set of gauge invariant operators in this theory has been constructed
in \cite{9202015} by using the open/closed string vertex that
emerges from the studies of \cite{thorn}. In this section we will
describe in detail these gauge invariant operators in OSFT and show
how they give rise to gauge invariant operators in VSFT.
It would be interesting to analyze the correlation functions of
these operators around the tachyon vacuum by using OSFT
in the  level truncation scheme. 

The same gauge invariant operators discussed here
have been considered independently by Hashimoto and Itzhaki,
who examined  the gauge invariance in an explicit 
oscillator construction, and motivated their role mostly in 
the context of OSFT \cite{0111092}.

We shall begin by reviewing the construction of ref.\cite{9202015} and
then we will consider the
generalization to VSFT.

\subsection{Gauge invariant operators in OSFT} 

The original cubic open string field theory \cite{OSFT} describing the
dynamics of the unstable D-brane, is described by the action:
\be \label{eg12}
S = -{1\over g_o^2} \,\Big[ {1\over 2} \langle \Phi, Q_B \Phi\rangle +
{1\over 3} \langle \Phi, \Phi * \Phi \rangle \Big]\, ,
\end{equation}
with gauge invariance:
\be \label{eg13}
\delta|\Phi\rangle = Q_B |\Lambda\rangle + |\Phi*\Lambda\rangle -
|\Lambda * \Phi\rangle\, .
\end{equation}
Here $Q_B$ is the BRST charge, $g_o$ is the open string
coupling constant, $|\Phi\rangle$ is the string field, and
$|\Lambda\rangle$ is the gauge transformation parameter. In this theory
there are gauge invariant operators $\OO_V(\Phi)$
corresponding to every on-shell closed string state represented by the
BRST invariant, dimension $(0,0)$ vertex operator $V =c \bar c V_m$,
where
$V_m$ is a  dimension
(1,1) primary in the bulk matter CFT. Given
any such closed string
vertex operator $V$,
we define $\OO_V(\Phi)$ as
the following linear function
of the {\it open}
string field $\Phi$:
\be \label{eg14}
\OO_V(\Phi) \equiv \langle h_1\circ (V(i) \Phi(0)) \rangle_{D}
= \langle V(0) h_1\circ \Phi(0) \rangle_{D} \,,
\end{equation}
where $h_N$ has been defined
in eq.\refb{eg15},
and $\langle ~\rangle_{D}$ denotes correlation
function on a unit disk.  Since $V$ is dimension
$(0,0)$ it is not affected by the conformal map $h_1$
despite being located at the singular point $z=i$.\footnote{
 In dealing
separately with ghost and matter contributions, however,
it may be useful to define
$\OO_V(\Phi)$ as $\lim_{\epsilon\to 0^+}\lim_{\eta\to
0^+}
\langle V(-\eta) h_{1+\epsilon}\circ \Phi(0)\ra$.
}
In \cite{9202015} these operators were added to the OSFT
action and it was shown that the resulting Feynman rules would
generate a cover of the moduli spaces of closed Riemann surfaces
{\it with boundaries} and closed string punctures
thus producing the appropriate closed
string amplitudes. 
The operators $\OO_V(\Phi)$  
can be
interpreted as the open string one point function
\be \label{eg16}
\OO_V(\Phi)= \langle \II | V(i) |\Phi\rangle\, ,  
\end{equation}
where $\langle \II |$ is the identity state of the $*$-product.
The world sheet picture is clear, 
$\OO_V(\Phi)$ 
corresponds to the amputated version of a semi-infinite strip
whose edge represents an open string, the two halves of which
are glued and a closed string vertex operator is located at the
conical singularity. 
Gauge
invariance of $\OO_V(\Phi)$
under \refb{eg13}  follows from the BRST invariance
of $V$ and the relations
\be \label{eg17}
|A\rangle * ( V(i) |B\rangle) =
V(i) |A * B\rangle, \qquad
( V(i) |A\rangle) * |B\rangle = V(i) |A * B\rangle\,.
\end{equation}

\subsection{Gauge invariant operators in VSFT}

Since the VSFT  field $\Psi$
must be related
to the original  unstable D-brane OSFT field $|\Phi\rangle$
by a field redefinition, the existence of gauge invariant
observables
in the OSFT implies that there must exist
such quantities in
the VSFT
as well. Even though the explicit relation between
$|\Psi\rangle$ and $|\Phi\rangle$ is not yet known,
we now argue that the VSFT gauge invariant
observables actually take the
same form as in OSFT.

The possible field redefinitions relating VSFT and OSFT
were discussed in ref.\cite{0012251}. If we denote by
$|\Phi_0\rangle$ the classical OSFT
solution describing the tachyon
vacuum, then
the shifted string field $|\wt\Phi\rangle
=|\Phi\rangle
-|\Phi_0\rangle$
may be related to $|\Psi\rangle$
by homogeneous redefinitions preserving the
structure of the cubic vertex, namely
\be \label{eg18}
(g_o^2 \kappa_0)^{1/3}\, |\Psi\rangle = e^{-K} |\wt\Phi\rangle\, ,
\end{equation}
where $K$ satisfies:
\ben \label{eg19}
K(A*B) &=& (KA) * B + A * (KB)\, , \nonumber \\
\langle KA, B\rangle &=& - \langle A, K B\rangle\, .
\end{eqnarray}
The explicit normalization factor $(g_o^2 \kappa_0)^{1/3}$ on the left hand
side of eq.\refb{eg18} has been chosen to ensure the matching of the
cubic
terms in \refb{eg1} and \refb{eg12} (see eq.\refb{egx1}).
Two general class of examples of $K$ satisfying \refb{eg19} are:
\be \label{eg20}
K | A\rangle = \sum_n a_n K_n |A\rangle\, ,
\end{equation}
where $K_n=L_n - (-1)^n L_{-n}$, and
\be \label{eg21}
K | A\rangle = |S * A\rangle - |A * S\rangle\, ,
\end{equation}
for some ghost number zero state $|S\rangle$.
Let us now consider the gauge invariant operator
\be \label{eg22}
\OO_V(\wt\Phi)
= \langle V(0) h_1\circ \wt\Phi(0)) \rangle_{D} = \langle \II | V(i)
|\wt\Phi\rangle\, ,
\end{equation}
invariant under the gauge transformation
\be \label{enewgauge}
\delta|\wt\Phi\rangle = Q_B |\Lambda\rangle +
|(\wt\Phi+\Phi_0)*\Lambda\rangle -
|\Lambda * (\wt\Phi+\Phi_0)\rangle\, ,
\end{equation}
and study what happens to this under an infinitesimal field redefinition
generated by a $K$ of the form \refb{eg20} or \refb{eg21}. It is easy to
see that both these field redefinitions preserve the form of $\OO_V$,
replacing $\wt\Phi$ by $(g_o^2 \kappa_0)^{1/3}\Psi$. For transformations of
the
form \refb{eg20}
this follows because $V(i)$, being a dimension zero primary,
commutes with the $K_n$'s and  the identity is annihilated by $K_n$.
For transformations of the form \refb{eg21}, form invariance follows
from eq.\refb{eg17}. Thus if $\wt\Phi$ and $\Psi$ are related by a field
redefinition of the form \refb{eg18}, with $K$ being a combination
of transformations of the type \refb{eg20} or \refb{eg21}, then we can
conclude that $\OO_V(\wt\Phi)$
is given by $(g_o^2 \kappa_0)^{1/3} \OO_V(\Psi)$, with
\be \label{eg23}
\OO_V(\Psi)
= \langle V(0) h_1\circ \Psi(0)) \rangle_{D} = \langle \II | V(i)
|\Psi\rangle\, .
\end{equation}

This must be a gauge invariant operator in VSFT.
Invariance of \refb{eg23} under the VSFT gauge transformation 
\refb{eg1ab} follows directly from \refb{eg17}, and the
relation $\langle\II|A*B\rangle = \langle A|B\rangle$.
Invariance under
\refb{eg1aa}
requires
\be \label{eg24}
\la \II | V(i) \QQ|\Lambda\ra \equiv
\langle h_1\circ (V(i) \QQ \Lambda(0)) \rangle_{D}=0\, .
\end{equation}
If we choose $\QQ$ to be of the form $\sum_n u_n \CC_n$, then for any
choice of the coefficients $u_n$, $\QQ$ commutes with $V(i)$. Thus if we
further restrict the $u_n$'s so that $\QQ$ annihilates $|\II\rangle$,
then
the gauge invariance of $\OO_V$ is manifest. Our choice
$\QQ=(c(i)-c(-i))/2i$, however, does not annihilate $|\II\rangle$ unless
we
define
$\QQ$ in a specific manner discussed in \refb{ep1a}.
Nevertheless, as we shall now show,
$\QQ$ annihilates $V(i)|\II\rangle$ independently of the definition
\refb{ep1a} and simply because of the collision of local ghost
insertions.  
Consider
a definition
of $\QQ$ that does not annihilate
$|\II\rangle$, by putting the operators in $\QQ$ at
$i+\epsilon$ for some finite $\epsilon$ and
then take the $\epsilon\to 0$
limit.
This gives:  
\ben \label{eg24a} 
&& \langle h_1\circ (V(i) \QQ \Lambda(0)) \rangle_{D}
\propto \lim_{\epsilon\to 0} \, \Bigl\langle h_1\circ \Big(V(i)
(c(i+\epsilon)-\bar c(i+\epsilon))
\Lambda(0)\Big)
\Bigr\rangle_{D} \nonumber \\
&=& \lim_{\epsilon\to 0}\,
\Bigl\langle V(h_1(i)) \Bigl\{  
{c(h_1(i+\epsilon))\over h_1'(i+\epsilon)}
- {\bar c(h_1(i+\epsilon))\over \bar h_1'(i+\epsilon))}
 \Bigr\} \,h_1\circ\Lambda(0)
\Bigr\rangle_{D}
\, . 
\end{eqnarray}
Using the results:
\be \label{eg25}
h_1(i+\epsilon) \sim \epsilon^2, \quad
h_1'(i+\epsilon) \sim \epsilon, \quad V(0) c(\eta)\sim \eta,
\qquad V(0) \bar c(\eta)\sim \eta\, ,
\end{equation}
we see that the expression between $\Bigl\langle \cdots \Bigr\rangle$
vanishes linearly in $\epsilon$.  
Thus $\OO_V(\Psi)$ defined
in \refb{eg23} is invariant under each of the transformations
\refb{eg1aa} and
\refb{eg1ab} for $\QQ$ given in \refb{ep1}.

\medskip  
It is interesting to relate the present discussion
to our observations on the cohomology of $\QQ$ below
equation \refb{ep1}. It was noted there that
$\QQ$ closed states had to have ghost insertions at the
open string midpoint. The question that emerges is whether
or not the gauge invariant operators discussed here are
$\QQ$ trivial. Presumably they are not. Indeed, thinking
of $c\bar c V_m$ as $c$ acting on $\bar c V_m$ we find that
the insertion of $\bar c V_m$, which is not of dimension zero
but rather of negative dimension, on a point with a
defect angle leads to a divergence. Therefore one cannot
think of the gauge invariant operators as ordinary trivial
states. Alternatively, one may wonder if the condition that
the closed string vertex operator $V$ be a dimension-zero primary
can be relaxed and still have $\OO_V$
be a sensible gauge invariant operator. Again, the answer is
expected to be no. Inserting an operator 
with dimension different from zero at
the conical singular point either gives zero or infinity. Moreover, if the
operator is not primary there are also difficulties with equation
\refb{eg17}. 

\subsection{Classical expectation value of $\OO_V$} 

Given a classical solution
of VSFT representing a D-brane we can ask  what is the value of
\be \label{ev1}
\OO_V(\Psi_{cl}) = \la \II|V(i)|\Psi_{cl}\ra\, .
\end{equation}
For $|\Psi_{cl}\ra=|\Psi_g\ra \otimes |\Psi_m\ra$, and $V=c\bar c V_m$,
we
have:
\be \label{ev2}
\OO_V(\Psi_{cl})=\la \II_g|c\bar c(i)|\Psi_g\ra \, \la
\II_m|V_m(i)|\Psi_m\ra\, .
\end{equation}
The ghost factor is universal, $-$ common to all D-brane solutions, and
all closed string vertex operators of the form $c\bar c V_m$. If we take
$|\Psi_m\ra$ to be a solution of the form discussed in \cite{0106010},
representing a D-brane associated with some boundary CFT
$\wt{BCFT}$, 
then it
is easy to show following the techniques of \cite{0106010} that
$\la\II_m|V(i)|\Psi_m\ra$ has the interpretation of a one point
correlation
function on the disk, with closed string vertex opertor $V_m$ inserted
at the center of the disk, and the boundary condition associated with
$\wt{BCFT}$ on the boundary of the disk. This, in turn can be interpreted
as
$\la \wt B_m|V_m\ra$ where $\la \wt B_m|$
is the matter part of the boundary
state
associated with $\wt{BCFT}$ and $|V_m\rangle$ is the closed string state
created by the vertex operator $V_m$.

\section{Closed string amplitudes in VSFT}

In this section we give our proposal for the emergence
of pure closed string amplitudes in the context of VSFT.
The basic idea is that the open string correlation of
the gauge invariant observables discussed in the previous
section give rise to closed string amplitudes obtained
by integration over the moduli spaces of Riemann surfaces
{\it without boundaries}. In order to justify this we will
have to make use of the regularized version of VSFT.

\subsection{Computation of correlation functions of $\OO_V$}

\begin{figure}[!ht]
\leavevmode
\begin{center}
\epsfysize=5cm
\epsfbox{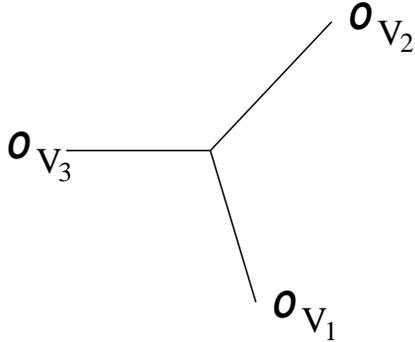}
\end{center}
\caption{The feynman diagram contributing to the correlation function
$\langle\langle \OO_{V_1} \OO_{V_2} \OO_{V_3}\rangle\rangle$.}
\label{f2}
\end{figure}
We shall now study correlation functions of the operators $\OO_V$
in VSFT.
In particular, we
shall analyze the following gauge
invariant correlation functions: 
\be \label{e5}
\lll  \OO_{V_1}(\Psi) \cdots \OO_{V_n}(\Psi)
\rrr
\end{equation}
where $\lll ~ \rrr$ stands for correlation functions in string field
theory and should not be confused with correlation functions in
two dimensional conformal
field theory. These correlation functions are calculated by the usual
Feynman rules of string field theory, $-$ in particular for $n=3$ the
tree
level correlation function receives contribution from just one Feynman
diagram shown in Fig.\ref{f2}. In computing these Feynman diagrams we
shall work with the regulated action \refb{eg28cc} and take the $a\to
0$
limit at the end. Including all the normalization factors, the Siegel
gauge
propagator is given by:
\be \label{enormprop}
a (\kappa_0(a))^{-1} {b_0\over L_0 + a} = {a \over \kappa_0(a)}
 b_0 \int_0^\infty
dl e^{-l (L_0+a)}
\, .
\end{equation}
We should, however, keep in mind
that this regularization procedure is ad hoc, and so the results obtained
from this should be interpreted with caution 
The correct regularization procedure presumably comes from replacing
the singular reparametrization discussed in
section \ref{s2origin} by a nearly singular reparametrization.

Since the propagator \refb{enormprop} is closely related to the
propagator of OSFT, and reduces to it up to an overall
normalization in the $a\to 0$ limit,
it will be useful to first review the calculation of
these
correlation function in OSFT around the D-25-brane background.
In
OSFT,
the
Feynman diagrams just have
closed string vertex operators
attached to strips of length $\ell_i$ and these
strips, together with internal open string propagators, 
are glued with three open
string vertices. So a typical diagram will have schematically
$$\prod_i \int_0^\infty d \ell_i  e^{-\ell_i L_0}
\prod_J \int_0^\infty d \ell_J  e^{-\ell_J L_0} $$
where the $\ell_J$ are intermediate propagator lengths. For an amplitude
with $n$ external closed strings there are altogether $(2n-3)$
propagators and
$(n-2)$ vertices.
Let us denote by $l_\alpha$ ($1\le \alpha\le (2n-3)$) the
lengths of the strips associated with these $(2n-3)$ propagators.
Thus the contribution to the amplitude can be
written as (ignoring powers of the open string coupling constant $g_o$):
\be \label{e52}
\int \prod_{\alpha=1}^{2n-3} dl_\alpha F(l_1, \ldots l_{2n-3})
\end{equation}
for some appropriate integrand $F$ which is computed in terms of
correlators of closed string vertex operators and ghost factors
associated
with the propagators on an appropriate Riemann surface.

If we repeat the
calculation in VSFT with the regularized propagator \refb{enormprop}, we
get
an additional
factor $e^{-a\sum l_\alpha}$ in the integrand.
This, in effect will restrict the integration
region to $l_\alpha$ of order $a^{-1}$ or less. Also each propagator
carries a multiplicative factor of $a/ \kappa_0(a)$
and each vertex carries a
multiplicative factor of $\kappa_0(a)$.
Thus the amplitude now takes the form:
\ben \label{e53}
A_n &=& (a/\kappa_0(a))^{2n-3}  (\kappa_0(a))^{n-2}
\int \prod_{\alpha=1}^{2n-3}
dl_\alpha
e^{-a\sum l_\alpha} F(l_1,
\ldots
l_{2n-3}) \nonumber \\
&=& \hskip-10pt (a^2/\kappa_0(a))^n a^{-3} \kappa_0(a) \int_0^\infty 
\hskip-10pt dv \int
\prod_{\alpha=1}^{2n-3}
dl_\alpha
\delta(v - \sum
l_\alpha) e^{-av}  F(l_1, \ldots
l_{2n-3})\, . \nonumber \\
\end{eqnarray}
We can absorb the $n$ factors of $(a^2/\kappa_0(a))$ into a multiplicative
renormalization of the operators $\OO_V$.
Using eq.\refb{es42} with  $\EE_a=\TT_{25}$,
the renormalized amplitude may be written as:
\be \label{e54}
A_n = {\TT_{25} \over a^3 F(a)} \int_0^\infty dv\, e^{-av}\, \int
\prod_{\alpha=1}^{2n-3} dl_\alpha \,\delta(v - \sum
l_\alpha)
F\Big(l_1, \ldots l_{2n-3}\Big) \, .
\end{equation}

\begin{figure}[!ht]
\leavevmode
\begin{center}
\epsfysize=8cm
\epsfbox{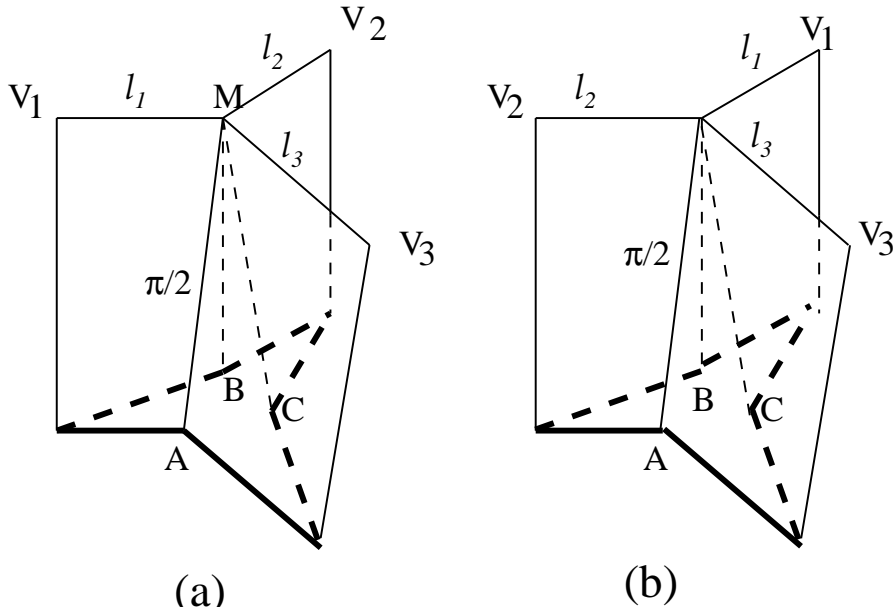} 
\end{center}
\caption{The Riemann surface representations of the Feynman diagram
contribution to the three closed string amplitude. $V_i$
denote the locations of the closed string vertex operators, $l_\alpha$
denote
the lengths of the strips representing open string propagators, and AMB,
BMC and CMA denote the three strings interacting via the three string
vertex with a common mid-point $M$. The thick line at the bottom is the
boundary of
the world-sheet diagram created from the Feynman diagram. The two diagrams
originate from two different contributions to the three string vertex,
corresponding to $\la A, \, B*C\ra$ and $\la A, \, C*B\ra$ respectively.
}
\label{f3} \end{figure}

$F(l_1, \ldots l_{2n-3})$
is computed by evaluating a
correlation function on a Riemann surface of the form shown in
Fig.\ref{f3}. Since $v$ in the above integral represents the sum of
the length parameters $l_\alpha$, we have
$l_\alpha\le v$, and
the closed string vertex operators are inserted within
a distance of order
$v$ of each other.
The boundary, shown by the thick line at the bottom, 
has length $2v$ since each length parameter
$l_\alpha$ contributes  
a length $2l_\alpha$ to the boundary. Finally, the
height of the diagram, measured
by the distance between the boundary and the closed string vertex
operators, 
is constant and
equal to $\pi/2$ -- this is because
open string strips have width $\pi$.
In addition to
the closed string vertex operators, the
correlator also includes an insertion of $b_0$ on each
propagator.

Let us
now rescale the metric
on this world-sheet by multiplying all
lengths by $\pi v^{-1}$.
In the resulting metric, and with $v$ now small, the Riemann surface
 looks
like a
long cigar of circumference $2\pi$ and height
$l_c=\pi^2/(2v)$.
All the
closed string vertex operators are inserted
within a finite distance of each other at the closed end of the cigar,
and their positions are naturally parametrized by quantities $u_\alpha$
defined, for $\alpha= 1,2\cdots,2n-3$, as
\be \label{e55}
u_\alpha  = 2\pi \,{l_\alpha \over v}
\quad
\to \quad \sum_\alpha u_\alpha = 2\pi\,.
\end{equation}
The other end of the cigar is open and represents the boundary of the
surface. The integration contours for
the $b$-integrals run parallel to the length of the cigar.
We will call this surface $\CC_v(\vec u)$, and as defined it is
a cylinder of height $\pi^2/(2v)$,
circumference $2\pi$, with one end open and the other sealed
and having closed string punctures with positions parameterized
by the $u_\alpha$.
We can use $v$ and $u_\alpha$ as independent variables of integration.
Since the $b$-contour integrals in the correlation function guarantee
that the integrand transforms as a volume form $dv \wedge du_1\wedge
\cdots$ in the moduli space, we can
formally denote these insertions as $B_v B_{\vec u}$, where
$B_v$ denotes
a single $b$ insertion associated with the  $v$-integration and $B_{\vec
u}$ is product of
$(2n-4)$ $b$-insertions associated with the integration
over $\vec u$.
Calling $\MM (\vec u)$ the
moduli space of $u_\alpha$'s,  the amplitude in
\refb{e54} can thus be written as
\be \label{e54a}
A_n = {\TT_{25} \over a^3 F(a)} \int_0^\infty dv\, e^{-av}\,
\int_{\MM(\vec u)}
\langle V_1\cdots V_n B_v \Bu \rangle_{\CC_v(\vec u)} \, .
\end{equation}

\begin{figure}[!ht]
\leavevmode
\begin{center}
\epsfysize=8cm
\epsfbox{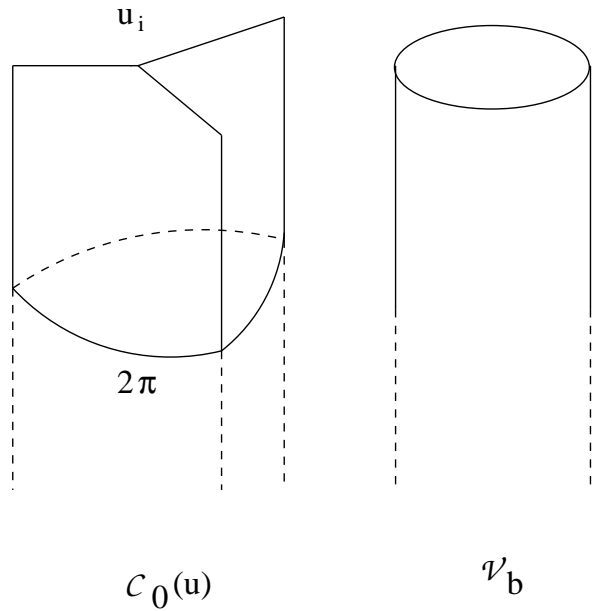}
\end{center}
\caption{Pictorial representation of $\CC_0(\vec u)$ and $\VV_b$ which are
glued together to produce the surface $\CC_v(\vec u)$.
}
\label{f5} \end{figure}

In order to proceed further we build the surface $\CC_v(\vec u)$
by sewing the semi-infinite cylinder $\CC_0(\vec u)$, obtained when
$v=0$, to the closed/boundary vertex $\VV_b$ represented by a
semi-infinite
cylinder of circumference $2\pi$ ending on an open boundary.
If we denote
by $w_1$ and $w_2$ the coordinates
used to describe the above two cylinders $\CC_v(\vec u)$
and $\VV_b$,
with $w_i
\equiv w_i + 2\pi$ and $\Im (w_i)<0$, 
and we let $z_i = \exp
(-iw_i)$; the sewing relation $z_1z_2 = t$, with real $t$ produces the
surface $\CC_v$ with $v = -\pi^2/(2\ln t)$. 
We therefore have that the
amplitude in question can be written as:
\ben \label{e54b}
A_n &=& {\TT_{25} \over a^3 F(a)} \int_0^\infty dv\, e^{-av}\,
\int_{\MM(\vec u)} \sum_k
\langle V_1\cdots V_n \Bu \chi_k \rangle_{\CC_0(\vec u)}\nonumber \\
&& \qquad\qquad \qquad\qquad\qquad\quad \cdot\,\, \langle \chi_k^c| B_v
e^{-{\pi^2\over 2v}
(L_0 + \bar
L_0)} |\VV_b
\rangle\, ,
\end{eqnarray}
where the $\chi_k$ is a basis element in the
space of ghost number two closed string vertex
operators, $\chi_k^c$ is the conjugate basis of ghost number four vertex
operators satisfying $\langle \chi_k^c|\chi_l\rangle=\delta_{kl}$,
$|\VV_b\rangle$ denotes the boundary state associated with the D-brane
under consideration, and
$L_0, \bar L_0$ refers to the closed string Virasoro generators.
In the first correlator, ${\chi_k}$ is inserted on the
puncture at infinity, and the second correlator is the one
point function on the semi-infinite cylinder.

We now need to determine $B_v$. This is done by going to the
$l_c=\pi^2/2v$ coordinate system, and using the transformation property
of
the $b$-insertions under a change of coordinates. In particular, we have
\be \label{ecoorch}
B_v dv = B_{l_c} d l_c\, .
\end{equation}
Furthermore the form of $B_{l_c}$ is well known, $-$ it simply
corresponds
to an insertion of a contour integral of $(b+\bar b)$ along the
circumference of the cigar. We shall denote this by $(b_0+\bar b_0)$.
This gives:
\be \label{ebv}
dv B_v = {\pi^2\over 2 v^2} dv (b_0+\bar b_0)\, .
\end{equation}
Substituting this into eq.\refb{e54b} we get,
\ben \label{e54bn}
A_n  &=&
{\TT_{25} \over a^3 F(a)}\, {\pi^2\over 2}\,
\int_0^\infty
{dv\over v^2} \, e^{-av}\,
\int_{\MM(\vec u)} \sum_k
\langle V_1\cdots V_n \Bu \chi_k \rangle_{\CC_0(\vec u)}\nonumber \\
&& \hskip60pt \cdot\,  \langle \chi_k^c| (b_0+\bar b_0) e^{-{\pi^2\over 2v}
(L_0 +
\bar L_0)} |\VV_b
\rangle\, ,
\end{eqnarray}

The key geometrical insight now is that
the moduli space $\MM (\vec u)$
defines a space of surfaces
$\CC_0(\vec u)$ which is precisely the moduli space $\MM_{n+1}$
of $n+1$-punctured
spheres.  This is a rigorous result and follows
from  a new minimal area problem that will be discussed in the next
subsection.  Therefore the integral above can be written as
\ben
\label{econt}  
&&
{\TT_{25} \over a^3 F(a)}\, {\pi^2\over 2}\,
\sum_k \int{dv \over v^2} e^{-a v} e^{-{\pi^2\over 2 v} (h_k
+ \bar h_k)} \langle \chi_k^c | (b_0 + \bar b_0) |
\VV_b\rangle \nonumber \\
&& \hskip 60pt \cdot \,\int_{\MM_{n+1}} \langle
V_1\cdots V_n \Bu \chi_k \rangle_{\CC_0(\vec u)}
\nonumber\\
&=& \sum_k C_k A_c(V_1, \ldots V_n, \chi_k)\, ,
\end{eqnarray}
where $(h_k, \bar h_k)$ is the conformal weight of $\chi_k$, and
$A_c(V_1,\ldots V_n, \chi_k)$ is
the $(n+1)$-point closed string amplitude of states
$V_1,\ldots V_n$ and $\chi_k$. $C_k$ are constants defined as:
\be \label{edefck} 
C_k =
{\TT_{25} \over a^3 F(a)}\, {\pi^2\over 2}\,
\int{dv \over v^2} e^{-a v} e^{-{\pi^2\over 2 v} (h_k
+ \bar h_k)} \langle \chi_k^c | (b_0 + \bar b_0) |
\VV_b\rangle\, .
\end{equation}
The multiplicative factor $C_k$ is non-zero in the
$a\to 0$
limit only for $h_k+\bar h_k\le 0$. For this range of values of $(h_k,\bar
h_k)$ $C_k$'s are actually infinite due to the divergence in the
$v$-integral from $v\simeq 0$ region. However,
note that the multiplicative factor $\TT_{25}/(a^3 F(a))$ vanishes as
$a\to\infty$ as is seen from Fig.\ref{f1}.
Thus this competes against the
divergent $v$-integral.  It will be interesting to see if in the correct
regularization procedure inherited from OSFT, the divergences in the $v$
integral are also regulated (as will happen, for example, if the kinetic
operator is multiplied by an additional factor of $e^{\epsilon L_0}$ for
some small $\epsilon$), and the final answer for the closed string
amplitude is actually finite.

We also note that among the contributions to \refb{econt} is the
contribution due to the zero momentum dilaton intermediate state.
By the soft dilaton theorem,
this is proportional to the on-shell $n$-point closed string amplitude on
the sphere. One could again speculate that in the correct regularization
procedure this is the only contribution that survives, and so the
correlation function \refb{e5} in the correctly regularized VSFT actually
gives us back the on-shell $n$-point amplitude at genus zero. A similar
argument has been given in \cite{closedbsft} in the context of boundary
string
field theory.

Since the regularization procedure we have been using is ad hoc, one can
ask what aspect of our results can be trusted in a regularization
independent manner. To this end, note that if the kinetic operator is
simply $c_0$, then the corresponding propagator is represented by a strip
of zero length. Thus
whatever be the correct regularization procedure, the regulated
propagator will
be associated with strips of small lengths if the regularization parameter
(analog of $a^{-1}$) is small. As our analysis shows, in this case the
corresponding Feynman diagram contribution to \refb{e5} will be associated
with a world-sheet diagram with small hole, and this, in turn, is related
to genus zero correlation functions of
closed string vertex operators 
with one additional closed string
insertion. Thus we can expect that whatever be the correct regularization
procedure, the correlation function \refb{e5} will always be expressed in
terms of a genus zero correlation function of closed string
vertex operators.

In the absence of a proper understanding of the correct regularization
procedure of the VSFT propagator, a
more direct approach to the problem of computing closed string
amplitude
in the tachyon vacuum will be to
try to do this computation directly
in OSFT around the tachyon vacuum. There are two competing
effects. On the one hand we have divergence due to the dilaton and other
tadpoles. On the other hand, the
coefficient of the divergence vanishes since the tachyon vacuum has zero
energy. Both of these are regulated in level truncation. Thus it is
conceivable that if we compute the correlation functions of
the operators
$\OO_V$ in OSFT around the tachyon vacuum by first truncating the theory
at
a given level $L$, and
then take the limit $L\to\infty$, then we shall get a finite result for
these correlation functions.

\subsection{Closed string moduli from open string moduli} \label{s61}

We have seen in the previous subsection that the calculation
of a correlator of gauge invariant
observables in regulated VSFT
can be related to the amplitude involving
closed string states 
parametrizing these observables
if a certain kind of string diagrams produces a full cover of the
moduli space of closed Riemann surfaces with
punctures. 
The diagrams in question are obtained by drawing all the diagrams
of OSFT supplemented by the open/closed vertex with the constraint
that the total boundary length is $2v$. Here 
$v = \sum l_\alpha$ where the $l_\alpha$'s  
are the lengths of the open string propagators. The diagrams 
are
then conventionally scaled to have cylinder with a total boundary length
of
$2\pi$ and  height of $\pi^2/(2v)$. The patterns of gluing are described
by the parameters $u_\alpha \geq 0$ defined in \refb{e55} and satisfying
$\sum_\alpha u_\alpha = 2\pi$.  At this stage one lets $v\to 0$ and thus
the cylinder becomes semi-infinite, with the boundary turning into
the $(n+1)$-th puncture.  The claim is  that the set of surfaces obtained
by letting the $u_\alpha$ parameters vary generate precisely the moduli
space of $(n+1)$ punctured spheres.

\medskip
In order to prove this we will show that the above diagrams
arise as the solution of a minimal area problem. As is well-known,
minimal area problems guarantee that OSFT, closed SFT, and
open/closed SFT generate
full covers of the relevant moduli spaces.\footnote{In the case of
OSFT, the first proof of cover of moduli space was given in
\cite{gmw} who focused on the case of surfaces without open string
punctures, and argued that by factorization the result extends
to the case with punctures. In \cite{Zwiebach:1991az} 
a direct proof based on
minimal area metrics is seen to apply for all situations.} The
basic idea is quite simple; given a specific surface, 
the  metric of minimal
area under a set of length conditions exists and is unique. 
Thus if we can establish a one to one correspondence between the string
diagrams labelled by $\{u_\alpha\}$ and such metrics, we would establish
that the $u_\alpha$ integration region covers the moduli space in a one to
one fashion.
The minimal area problem for our present
purposes is the following

\medskip
\noindent
{\it Consider a genus zero
Riemann surface with $(n+1)$ punctures.
Pick  one special
pucture $P_0$, and find the minimal area metric under the
condition  that all curves homotopic
to $P_0$ have length larger or equal to $2\pi$.}

\medskip
As usual homotopy equivalence does not include moving curves
across punctures, thus a curve surrounding $P_0$ and $P_1$ is
not said to be homotopic to
a curve surrounding $P_0$.
This problem is a modification
of the minimal area problem defining the polyhedra of classical closed
string field theory \cite{9206084} -- in this case one demands that the
curves
homotopic to all the punctures be longer than or equal to $2\pi$
\cite{Saadi:1989tb}.

We use the principle of saturating geodesics to elucidate
the character of the minimal area metric solving our stated
problem. This principle
\cite{Zwiebach:1990ew} states that through every point in the string
diagram
there must exist a curve saturating the length condition.
Therefore the solution must take the form of a semi-infinite
cylinder of circumference $2\pi$. The infinite end represents
the puncture $P_0$. The other side must
be sealed somehow, and the other $n$ punctures must
be located somewhere in this cylinder.
Since there are no
length conditions for
the other punctures, they do not generate their own cylinders.

\smallskip
\noindent
Assume now that the other punctures are met successively as
we move up the cylinder towards the sealed edge.
This is actually impossible, as we now show. Let $P_1$
be the first puncture we meet as we move up from $P_0$.
Consider a saturating circle just below the first such puncture.
 That circle has
to be of length $2\pi $ since it is still homotopic to $P_0$.
If the cylinder continues to exist beyond $P_1$
a geodesic circle of length $2\pi$
just above $P_1$ is not homotopic any more to
$P_0$, and there is no length constraint on it anymore. This cannot
be a solution of the minimal area problem since the metric
could be shrunk along that circle without violating any
length condition.
This shows that all the punctures must be met at once.
Thus the picture is that of a semiinfinite cylinder, where
on the last circle the $n$ closed string punctures
are located, and the various segments of the circle
are glued to each other to seal the cylinder, so that
any nontrivial curve not homotopic to $P_0$ can be shrunk
to zero length.

This is exactly the pattern of the string diagrams
that we obtained. It is clear that the $u_\alpha$
parameters associated to a fixed Feynman graph
are in fact gluing parameters. Thus the string diagrams
solve the minimal area problem and
due to the uniqueness of the minimal area metric
they do not double
count.
Can they miss any surface ? There are two alternative ways to see
that the answer is no.  
First, the space of $u_\alpha$ parameters  has no
codimension one boundaries, and includes all the requisite degenerations of
the $(n+1)$ punctured sphere associated with the collision of two or more
punctures. Since these are the standard properties of moduli spaces,
no surfaces can be missing. Second,
for any surface there is a string diagram
-- this is guaranteed because this minimal area problem
is known to have a solution defined by a Jenkins-Strebel
quadratic differential. Such quadratic differential builds
a string diagram consistent with our Feynman rules, and thus
must have been included.

We illustrate the above result with an example, the case of
four-punctured spheres generated by considering the correlation
of three gauge invariant observables.
We shall explain that the only boundaries of the $u_\alpha$
integration region are the known boundaries of the moduli space
corresponding to degeneration of the four-punctured sphere.
In this case
three strips of lengths $\ell_1,\ell_2,\ell_3$ representing the three
external propagators
are joined by a 3 open string vertex -- no internal
propagator is possible
here.
The amplitude contains sum over two different world-sheet digrams,
coming from
two different cyclic
arrangements of the open strings at the vertex, as shown in
Fig.\ref{f3}(a) and (b).
If we
denote
by $l_1$,
$l_2$, $l_3$ the lengths of the strips associated with the open string
propagators, and $v = l_1+l_2 + l_3$, then the region of integration,
with
$u_i= 2\pi l_i/v$, is
\be \label{epara}
u_i \geq 0\,, \quad  \sum u_i = 2\pi\,.
\end{equation}
There are apparently three codimension one boundaries of the $u_i$
integration
region,
associated with each $u_i =0$. These correspond to $l_i=0$.
It is easy, however, to see from Fig.\ref{f3} 
that the configuration $l_i=0$ for any $i$ are actually identical
configurations in
the two diagrams, and hence in the sum of two diagrams the $l_i=0$
configuration simply marks the transition from the component of the
moduli
space covered by the first diagram to another 
component of the moduli space
covered by the second diagram. On the other hand the  codimension two
boundaries corresponding to the three cases of $u_i = 2\pi$,
represent the configurations where two length parameters vanish and
produce the expected degenerations of the 4-punctured sphere. In
particular the
$l_i=l_j=0$ configuration represents the degeneration where the $i$-th
and
the
$j$-th vertex operators come close to each other, and the other vertex
operator approaches, in the conformal sense, the boundary of the surface.

Indeed even if the height of the cylinders is finite we are producing
a boundaryless subspace of the moduli space of a sphere with three
punctures and one hole.  As the height of the cylinder goes to infinity
we really have four punctures and again we are producing a boundaryless
moduli space involving four punctures on a sphere and all the requisite
degenerations. This must be the moduli space of four punctured spheres.

The generalization to the case of $n$-point amplitude
is straightforward.
Any codimension one boundary
corresponding to a single $l_i$ vanishing marks a
transition to another component of the moduli space represented by
another
diagram, whereas if a group of $l_i$ associated with a connected
part
of the diagram, and
containing at least two external propagator vanishes,
it corresponds to a degeneration of the Riemann surface.
A detailed argument along the lines of \cite{Zwiebach:1991az} should
be possible to construct, but as we do not expect complications,
we shall not attempt to give the complete argument here.

\bigskip
The discussion above clearly holds for surfaces of arbitrary
genus, and the minimal area problem is just the same one.
More interestingly, however, the discussion also generalizes
for the case of multiple boundary components. Given our Feynman
rules of regularized VSFT, the analysis of the previous subsection
would lead to surfaces in which each boundary component would
give rise to a semi-infinite cylinder of circumference $2\pi$.
The various cylinders would join simultaneously with a generalized
set of $u_\alpha$ parameters describing their gluing.
If the Feynman graph represents a surface of genus $g$ with $n$
gauge invariant operators and $b$ boundaries, the space of
$u_\alpha$ parameters will generate the moduli space
${\cal M}_{g, n+b}$ of genus $g$ boundariless Riemann surfaces
with $n+b$ punctures. The associated minimal area problem justifying
this result would consider the metric of minimal area on a genus
$g$ surface with $n+b$ punctures under the condition that all curves
homotopic to the $b$ punctures be longer than or equal to $2\pi$.
Thus the correlation function would reduce to the pure closed string
amplitude
of $n$ closed string vertex operators and $b$ zero momentum massless
states.

\sectiono{Discussion}

In this paper we have fixed a specific form of
the kinetic term $\QQ$ of VSFT thus giving a precise
definition of the theory and making it possible to
study in detail various questions. While the selected
$\QQ$ is special in several ways, VSFT thus defined
needs regulation for some but not all computations.
Our regulation of VSFT is admittedly somewhat tentative.
If VSFT can be shown explicitly to arise as
a singular reparametrization of the OSFT action expanded
around the tachyon
vacuum,  a more natural  regulator may be obtained
by viewing the reparametrization as a flow and using
the representatives near the singular endpoint.

We believe other results presented in this paper may have uses beyond
the ones investigated presently.

\begin{itemize}

\item Our explicit
level expansion calculations have uncovered
the existance of surface states different from the sliver
and still satisfying the projector condition. These
new projectors may have important applications.

\item  The twisted CFT used 
to obtain exact analytic solutions may be
a useful tool to obtain exact solution of string field
theory even for the original OSFT representing
the vacuum around unstable D-branes.

\item We have uncovered local gauge invariant
operators in open string field theory. Their natural
relation to closed string vertex operators is reminiscent
of AdS/CFT, and of gauge invariant operators in non-commutative
gauge theory. There could be interesting uses for these
operators in studying observables of VSFT. 

\item We have seen
how closed string moduli arise from  the open string moduli of regulated
VSFT, by noting how a minimal area problem involving open string curves
naturally
dovetails into a minimal area problem involving closed string curves.
This, we believe may capture the essence of the mechanism by which closed
strings emerge in vacuum string field theory.

\end{itemize}

We end  this  section  by presenting
two possible modifications
of VSFT that could have intriguing applications. 

\begin{itemize}

\item It may possible to relax naturally the purely
ghost condition on the kinetic operator $\QQ$
while preserving the fact that $\QQ$ is a local
insertion.  Certain kinds of matter insertions
would not destroy the key properties that guaranteed
some of the successes of VSFT. For example, our present choice of
$\QQ$ -- the ghost field at the open string midpoint,
could perhaps be modified by
multiplication of operators involving the matter
stress tensor also inserted at (or near) the mid-point. It is not
clear if one could satisfy the conditions of nilpotency
of $\QQ$ and absence of cohomology, but if so, modified
slivers with stress tensor insertions at (or near) the midpoint
could yield solutions representing D-branes, and the computation of
ratios of tensions of D-branes  described in
\cite{0105168,0106010} would hold.

\item
It appears to be possible to formulate a version of VSFT
where the original sliver of the matter and $(b,c)$ system
is a solution, and the action for the sliver is finite.
This requires, however, using string fields of ghost number
zero, and the connection with the usual string field theories
that use ghost
number one string fields could reintroduce singular behavior.
Consider 
the operator $Y(z)={1\over 2}c\partial c
\partial^2 c(z)$, a dimension zero primary of ghost number three.
Being of dimension zero such an 
operator can be readily inserted at the
string
midpoint. We can therefore try $Y(i) + Y(-i)$ as kinetic term of the
action:  
\be \label{vsft0}
S  = \beta\bigg[ {1\over 2} \langle \Phi |(Y(i) + Y(-i))
| \Phi\rangle
+ {1 \over 3} \langle \Phi | (Y(i) + Y(-i)) |\Phi * \Phi \rangle \bigg]
\end{equation}
where now the string field $\Phi$ must be of ghost number zero, and
$\beta$ is a constant.
This action is invariant under the homogeneous gauge transformation
$\delta|\Phi\ra = |\Phi*\Lambda\ra - |\Lambda * \Phi\ra$. 
No Fock space state can be annihilated by $Y(i) + Y(-i)$ and therefore
there
are no conventional physical states.
The equation of motion  becomes $(Y(i) + Y(-i)) \Bigl( | \Phi\rangle
+   |\Phi * \Phi \rangle \Bigr) =0$ and (minus) the sliver provides a
simple solution. The value of the action at the solution is
$S\, = {1\over 6} \beta \langle \Xi | (Y(i) + Y(-i)) |
\Xi\rangle$.
Since the sliver is constructed as an exponential of Virasoro operators
of zero central charge acting on the vacuum, and $Y$ is a dimension zero
primary,
the value of the action is  $S\, = {1\over 6} \beta \langle 0 |
(Y(i) +
Y(-i)) | 0\rangle = {1\over 3} \beta$.
This can be made to 
agree with the expected answer by appropriate choice of the finite
constant $\beta$.

Although the relationship of the action \refb{vsft0} 
to the original open string field
theory is not clear, this action 
has the property that given
any boundary conformal field theory, we can construct, following
\cite{0105168,0106010}, a solution to the
equations of motion such that the energy density
associated
with the solution is equal to the tension of the corresponding D-brane.
This is one of the properties that must be satisfied by any string field
theory that represents string field theory around the tachyon vacuum.

\end{itemize}

\noindent{\bf Acknowledgements}:
We would like to thank  C.~Imbimbo, A.~Hashimoto, 
N.~Itzhaki, F.~Larsen, S.~Mathur, N.~Moeller,
P.~ Mukhopadhyay, H.~Ooguri,
M.~Schnabl, and W.~Taylor for useful discussions.
The work of L.R. was supported in part
by Princeton University
``Dicke Fellowship'' and by NSF grant 9802484.
The  research of A.S. was supported in part by a grant 
from the Eberly College 
of Science of the Penn State University.
The work of  B.Z. was supported in part
by DOE contract \#DE-FC02-94ER40818.

\end{document}